\documentclass[12pt,aps,preprint,nofootinbib,superscriptaddress,nobalancelastpage]{revtex4}
\pdfoutput=1

\usepackage{mathrsfs}
\usepackage{amsmath,amsthm,amssymb}
\usepackage{color}
\usepackage{graphicx}
\usepackage{verbatim}
\usepackage{multirow}

\usepackage{hyperref}
\hypersetup{colorlinks,bookmarksopen,bookmarksnumbered,citecolor=blue,
linkcolor=blue,pdfstartview=FitH,urlcolor=blue}
\usepackage{xspace}

\newcommand{\be}{\begin{equation}}
\newcommand{\ee}{\end{equation}}
\newcommand{\bea}{\begin{eqnarray}}
\newcommand{\eea}{\end{eqnarray}}

\newcommand{\eq}{Eq.}

\newcommand{\ie}{\emph{i.e.}}

\DeclareMathOperator{\Rea}{Re}
\DeclareMathOperator{\Ima}{Im}

\DeclareMathOperator{\diag}{diag}

\definecolor{cadmiumgreen}{rgb}{0.0, 0.65, 0.31}

\begin{document}
\preprint{IFT-UAM/CSIC-16-090, FTUAM-16-35, FERMILAB-PUB-16-400-T}

\title{Non-Unitarity, sterile neutrinos, and Non-Standard neutrino Interactions}

\author{Mattias Blennow}
\email{emb@kth.se}
\affiliation{Department of Theoretical Physics, School of Engineering Sciences, KTH Royal Institute of Technology, 
  Albanova University Center, 106 91 Stockholm, Sweden}
\author{Pilar Coloma}
\email{pcoloma@fnal.gov}
\affiliation{Theoretical Physics Department, Fermi National Accelerator Laboratory, \\
P.O. Box 500, Batavia, IL 60510, USA  }
\author{Enrique Fernandez-Martinez}
\email{enrique.fernandez-martinez@uam.es}
\affiliation{Departamento de F\'isica Te\'orica, Universidad Aut\'onoma de Madrid, Cantoblanco E-28049 Madrid, Spain}
\affiliation{Instituto de F\'isica Te\'orica UAM/CSIC,
 Calle Nicol\'as Cabrera 13-15, Cantoblanco E-28049 Madrid, Spain}
\author{Josu Hernandez-Garcia}
\email{josu.hernandez@uam.es}
\affiliation{Departamento de F\'isica Te\'orica, Universidad Aut\'onoma de Madrid, Cantoblanco E-28049 Madrid, Spain}
\affiliation{Instituto de F\'isica Te\'orica UAM/CSIC,
 Calle Nicol\'as Cabrera 13-15, Cantoblanco E-28049 Madrid, Spain}
\author{Jacobo Lopez-Pavon}
\email{jacobo.lopez.pavon@cern.ch}
\affiliation{INFN, Sezione di Genova, via Dodecaneso 33, 16146 Genova, Italy}
\affiliation{CERN, Theoretical Physics Department, Geneva, Switzerland.}

\begin{abstract}
{The simplest Standard Model extension to explain neutrino masses involves the addition of right-handed neutrinos. 
At some level, this extension will impact neutrino oscillation searches.
In this work we explore the differences and similarities between the case in which these 
neutrinos are kinematically accessible (sterile neutrinos) or not (mixing matrix non-unitarity). We clarify apparent
inconsistencies in the present literature when using different parametrizations to describe these effects and recast
both limits in the popular neutrino non-standard interaction (NSI) formalism. We find that, in the limit in which sterile 
oscillations are averaged out at the near detector, their effects at the far detector coincide with non-unitarity at leading 
order, even in presence of a matter potential. We also summarize the present bounds
existing in both limits and compare them with the expected sensitivities of near-future facilities taking the DUNE 
proposal as a benchmark. We conclude that non-unitarity effects are too constrained to impact present or near 
future neutrino oscillation facilities but that sterile neutrinos can play an important role at long baseline experiments. 
The role of the near detector is also discussed in detail. }
\end{abstract}

\maketitle

\section{Introduction}
\label{sec:intro}

The simplest extension of the Standard Model (SM) of particle physics able to account for the evidence for neutrino masses and mixings is the addition of right-handed neutrinos to its particle content. A Majorana mass term for these new singlet fermions is then allowed by all symmetries of the Lagrangian. This new mass scale at which lepton number is violated could provide a mechanism to also explain the origin of the observed matter-antimatter asymmetry of the Universe~\cite{Fukugita:1986hr} and is a necessary missing piece to solve the mysterious flavour puzzle. However, given that this new mass scale is not related to the electroweak symmetry breaking mechanism, there is no theoretical guidance for its value. A large Majorana scale leads to the celebrated seesaw mechanism~\cite{Minkowski:1977sc,Mohapatra:1979ia,Yanagida:1979as,GellMann:1980vs}, providing a very natural explanation of the lightness of neutrino masses. On the other hand, it also leads to unnaturally large contributions to the Higgs mass, worsening the hierarchy problem~\cite{Casas:2004gh}. Conversely, a light neutrino mass could also naturally stem from a symmetry argument~\cite{Mohapatra:1986bd,Bernabeu:1987gr,Branco:1988ex,Buchmuller:1990du,Pilaftsis:1991ug,Kersten:2007vk,Abada:2007ux}. Indeed, neutrino masses are protected by the lepton number symmetry and, if this is an approximate symmetry of the theory, a large hierarchy of scales is not required to naturally accommodate the lightness of neutrinos. Thus, the value of this scale of new physics can only be probed experimentally and, depending on its value, very different and interesting phenomenology would be induced in different observables. 

In this work we analyze the phenomenological impact of these new physics in neutrino oscillation facilities. If the new mass scale is kinematically accessible in meson decays, the sterile states will be produced in the neutrino beam. On the other hand, if the extra neutrinos are too heavy to be produced, the effective three by three PMNS matrix will show unitarity deviations. We will refer to these situations as sterile and non-unitary neutrino oscillations, respectively. The aim of this work is to discuss in which limits these two regimes lead to the same impact on the oscillation probabilities, and reconcile apparently inconsistent results in previous literature. 

This work is organized as follows. In section \ref{sec:comparison} we will compare the non-unitarity and sterile neutrino phenomenology and discuss in which cases both limits are equivalent.  In Section \ref{sec:param} we will present and solve the apparent inconsistency present in the literature studying non-unitarity effects in different parametrizations, and provide a mapping between the two. In Section \ref{sec:NSI} we will recast both scenarios using the popular NSI parametrization. The existing bounds, applicable in both regimes, from present observables are summarized in Section~\ref{sec:bounds}. Finally, in Section \ref{sec:simulations} we present the sensitivity of the prospective Deep Underground Neutrino Experiment (DUNE) to these effects, and our conclusions are summarized in Section \ref{sec:concl}.

\section{Non-unitarity and sterile neutrino phenomenology comparison}
\label{sec:comparison}

In this section we will show how, under certain conditions, the phenomenology of non-unitarity and sterile neutrino oscillations are equivalent to leading order in the active-heavy mixing parameters, not only in vacuum but also in matter. If $n$ extra right-handed neutrinos are added to the SM Lagrangian, the full mixing matrix (including both light and heavy states) can be written as
\begin{equation}
\label{eq:Ufull}
\mathcal{U} = \begin{pmatrix}
N & \Theta  \\
R & S \\
\end{pmatrix} \, ,
\end{equation}
where $N$ represents the $3 \times 3$ active-light sub-block (i.e., the PMNS matrix), which will no longer be unitary\footnote{Note that this is true regardless whether the extra states are kinematically accessible or not.}. Here, $\Theta$ is the $3\times n$ sub-block that includes the mixing between active and heavy states, while the $R$ and $S$ sub-blocks define the mixing of the sterile states with the light and heavy states, respectively. Note that both $R$ and $S$ are only defined up to an unphysical rotation of the sterile states and that neither of them will be involved when considering oscillations among active flavours.

\subsection{Non-unitarity case}
In the case of non-unitarity, only the light states are kinematically accessible and the amplitude for producing one of these states in conjunction with a charged lepton of flavour $\alpha$ in a particular decay is proportional to the mixing matrix element $N_{\alpha i}^*$. In the mass eigenstate basis, the evolution of the produced neutrino state is given by the Hamiltonian~\cite{Antusch:2006vwa}
\begin{equation}
\label{eq:Hnonunitarity}
H = \frac{1}{2E}\begin{pmatrix}
0 & 0 & 0 \\
0 & \Delta m_{21}^2 & 0 \\
0 & 0 & \Delta m_{31}^2
\end{pmatrix} + N^\dagger  \begin{pmatrix}
V_{\rm CC}+V_{\rm NC} & 0 & 0 \\
0 & V_{\rm NC} & 0 \\
0 & 0 & V_{\rm NC}
\end{pmatrix}  N,
\end{equation}
where $V_{\rm CC}=\sqrt{2}G_Fn_e$ and $V_{\rm NC}=-G_F n_n/\sqrt{2}$ are the charged-current (CC) and neutral-current (NC) matter potentials, respectively. The oscillation evolution matrix $S^0$ in this basis is now defined as the solution to the differential equation
\begin{equation}
i \dot S^0 = H S^0
\end{equation}
with the initial condition $S^0(0) = I$, $I$ being the identity matrix. For a constant matter potential, this equation has the formal solution
\begin{equation}
S^0 = \exp(- i H L).
\label{S0}
\end{equation}
The amplitude for a neutrino in the mass eigenstate $j$ to interact as a neutrino of flavour $\beta$ is given by the mixing 
matrix element $N_{\beta j}$, which means that the oscillation probability will be given by
\begin{equation}
\label{eq:nonunitarityoscillationformula}
P_{\alpha\beta} = |(N S^0 N^\dagger)_{\beta\alpha}|^2.
\end{equation}
Here $P_{\alpha\beta}$ denotes the ``theoretical'' oscillation probabilities (although it should be noted that they do not add up to one), defined as the ratio between the observed number of events divided by the product of the SM-predicted flux times cross section. In other words, $P_{\alpha\beta}$ is the factor that would be needed to obtain the number of events after convolution with the standard model flux and cross sections.

However, in practice neutrino oscillation experiments do not measure $P_{\alpha\beta}$. Most present and future experiments rather determine the flux and cross sections via near detector data and extrapolate to the far detector by correcting for the different geometries, angular apertures, and detection cross sections. In this scenario, the oscillation probability would then inferred from the ratio
\begin{equation}
\label{eq:experimentalprobabilitydefinition}
\mathcal P_{\alpha\beta} = \frac{R_\beta }{R_\alpha },
\end{equation}
where $R_\beta$ and $R_\alpha$ are the observed event rates at the far detector and the corresponding extrapolation of the near detector result, respectively. For the near detector, we assume that the phases corresponding to the propagation of the light neutrinos have not yet developed significantly and therefore $S^0 = I$, resulting in the experimentally inferred probability
\begin{equation}
\label{eq:experimentalprobability}
\mathcal P_{\alpha\beta} = \frac{\left| (N S^0 N^\dagger)_{\beta\alpha} \right|^2}{((NN^\dagger)_{\alpha\alpha})^2}.
\end{equation}
In the SM limit the matrix $N$ becomes unitary and, thus, $NN^\dagger=I$ and $\mathcal P_{\alpha\beta} = P_{\alpha\beta}$ as expected.

\subsection{Sterile neutrino case}

In the sterile neutrino scenario, all of the states are kinematically accessible and the full oscillation evolution matrix $\mathcal S$, involving both light and heavy states, takes the form
\begin{equation}
\mathcal S = \mathcal U \mathcal S^0 \mathcal U^\dagger,
\end{equation}
where $\mathcal S^0$ is the full $(3+n)\times (3+n)$ evolution matrix expressed in the mass eigenbasis. For vacuum oscillations, we find that $\mathcal S^0 = \diag(\exp(-i\Delta m_{j1}^2L/2E))$. Therefore, the active neutrino $3\times 3$ sub-block $S$ can be simplified to 
\begin{equation}
S_{\alpha\beta} = \sum_{i \in \rm light}  N_{\alpha i} S^0_{ij} N^*_{\beta j} + \sum_{J \in \rm heavy} \Theta_{\alpha J}\Theta_{\beta J}^* \Phi_J,
\end{equation}
where $\alpha,\beta$ stand for active neutrino flavors, $\Phi_J$ is the phase factor acquired by the heavy state $J$ as it propagates, and $S^0$ is defined in the same way as in the non-unitarity case. 

In the limit of large mass squared splitting between the light and heavy states (\ie, $\Delta m^2_{iJ} L/E \gg 1$) the oscillations are too fast to be resolved at the detector and only an averaged-out effect is observable. In this averaged-out limit, the cross terms between the first and second term in the evolution matrix average to zero and we find
\begin{equation}
\label{eq:sterileoscillationformula}
P_{\alpha\beta} = |S_{\alpha\beta}|^2 = \left| \sum_i N_{\alpha i} S^0_{ij} N^*_{\beta j} \right|^2 + \mathcal O(\Theta^4) \, ,
\end{equation}
which recovers the same expression as \eq~\eqref{eq:nonunitarityoscillationformula} up to the $\mathcal O(\Theta^4)$ corrections.\footnote{Note that this expression is also applicable whenever the light and heavy states decohere due to wave packet separation.} Thus, we can conclude that averaged-out sterile neutrino oscillations in vacuum are equivalent to non-unitarity to leading order (this equivalence is indeed lost at higher orders). We will therefore concentrate on this averaged-out limit for the rest of this paper.

For oscillations in the presence of matter, the sterile neutrino oscillations will be subjected to a matter potential that in the flavour basis takes the form
\begin{equation}
\mathcal H_{\rm mat}^f = \begin{pmatrix}
V_{3 \times 3} & 0 \\ 0 & 0
\end{pmatrix},
\end{equation}
where
\begin{equation}
V_{3\times 3} = \begin{pmatrix}
V_{\rm CC}+V_{\rm NC} & 0 & 0 \\
0 & V_{\rm NC} & 0 \\
0 & 0 & V_{\rm NC}
\end{pmatrix} \, .
\end{equation}
If the matter potential is small in comparison to the light-heavy energy splitting $\Delta m_{iJ}^2/2E$, the light-heavy mixing in matter will be given by
\begin{equation}
\tilde{\Theta}_{\alpha J} = \Theta_{\alpha J} + \frac{2E}{\Delta m_{iJ}^2}(\delta_{\alpha e} V_{\rm CC} \Theta_{eJ} + \Theta_{\alpha J} V_{\rm NC})
\end{equation}
to first order in perturbation theory. In the limit $\Delta m_{iJ}^2/2E \gg V_{\rm CC}, V_{\rm NC}$, we can therefore neglect the difference between the heavy mass eigenstates in vacuum and in matter, and apply Eq.~\eqref{eq:sterileoscillationformula}. Thus, we conclude from this that the matter Hamiltonian in the light sector can be computed in exactly the same way as for the non-unitarity scenario and we therefore find the very same expressions for the ``theoretical'' probability in Eq.~\eqref{eq:nonunitarityoscillationformula} as for the non-unitarity case, at leading order in $\Theta$.

In the case of sterile neutrinos one also needs to consider the impact of the near detector measurements on the extraction of the experimentally measurable probability. 
In this work we will always assume that the oscillations due to the additional heavy states are averaged out at the far 
detector. However, this might not be the case at the near detector. Ideally, both sets of observables should be 
simulated and analyzed together consistently. Nevertheless, the following simplified limiting cases can be identified:
\begin{enumerate}
\item The light-heavy oscillations are averaged out already at the near detector. For practical purposes, the oscillation phenomenology in this case is identical to the non-unitarity case and \eq~\eqref{eq:experimentalprobability} also applies. For the experimental setup of DUNE, that will be studied as an example of these effects in Section~\ref{sec:simulations}, with a peak neutrino energy of $\sim 2.5$~GeV and a near detector distance of $\sim0.5$~km, this is the case when $\Delta m^2 \gtrsim 100~\textrm{eV}^2$.
\item The light-heavy oscillations have not yet developed at the near detector, but are averaged out at the far detector. In this case, the near detector would measure the SM fluxes and cross sections, and therefore the denominator in Eq.~\eqref{eq:experimentalprobability} would be equal to one. In this case, the experimental probability would coincide with the ``theoretical'' probability in Eq.~\eqref{eq:nonunitarityoscillationformula}. This scenario is the one implicitly assumed in many phenomenological studies, given the simplicity of Eq.~\eqref{eq:nonunitarityoscillationformula}. However, it is typically only applicable in a very small part of the parameter space, i.e., for very particular values of $\Delta m^2$ (which depend on the neutrino energy and on the near and far detector baselines). For DUNE, since the far detector baseline is $1300$~km, this would be the case only in the region $0.1~\textrm{eV}^2 \lesssim \Delta m^2 < 1~\textrm{eV}^2$. This scenario will nevertheless be explored in Sec.~\ref{sec:simulations} to highlight its differences relative to the previous one, which is applicable in a larger fraction of the $\Delta m^2$ parameter space.
\item The oscillation frequency dictated by the light-heavy frequency matches the near detector distance. 
In this case, oscillations could be seen at the near detector as a function of neutrino energy, leading to more striking 
signals. At DUNE, this regime is matched for values of $\Delta m^2$ in the range presently favoured by the LSND/MiniBooNE~\cite{Aguilar:2001ty,Aguilar-Arevalo:2013pmq} and 
reactor anomalies~\cite{Mention:2011rk,Huber:2011wv} (see~\cite{Kopp:2013vaa,Giunti:2015wnd,Collin:2016rao} for recent reviews). This regime at DUNE has been already analized (see for instance Ref.~\cite{Choubey:2016fpi}). The sensitivity to this part of the parameter space will be dominated by the dedicated experiments built to explore these anomalies, such as the Fermilab short-baseline neutrino program~\cite{Antonello:2015lea}, leaving little room for their averaged-out effects to be observed at the far detectors in long-baseline oscillation experiments. Therefore, this scenario will not be discussed further.
\end{enumerate}

\section{Parametrizations}
\label{sec:param}

The two most widely used parametrizations to encode these non-unitarity effects stemming from the heavy-active mixing are 
\begin{equation}
N = (I-\eta) U'  \quad \mathrm{or} \quad N = T U = (I-\alpha) U ,
\label{eq:N}
\end{equation}
where $\eta$ is a Hermitian matrix~\cite{Broncano:2002rw,FernandezMartinez:2007ms} and $T$ is a lower triangular matrix~\cite{Xing:2007zj,Xing:2011ur,Escrihuela:2015wra,Li:2015oal}. In Eq.~\eqref{eq:N} both $U$ and $U'$ are unitary matrices that are equivalent to the standard PMNS matrix up to small corrections proportional to the deviations encoded in $\eta$ and $\alpha$. 
\begin{equation}
\eta = \begin{pmatrix}
\eta_{ee} & \eta_{e\mu} & \eta_{e\tau}\\
\eta_{e\mu}^* & \eta_{\mu\mu} & \eta_{\mu\tau}\\
\eta_{e\tau}^* & \eta_{\mu\tau}^* & \eta_{\tau\tau}
\end{pmatrix},
\quad 
\alpha = (1-T) = \begin{pmatrix}
\alpha_{ee} & 0 & 0\\
\alpha_{\mu e} & \alpha_{\mu \mu} & 0\\
\alpha_{\tau e} & \alpha_{\tau \mu} & \alpha_{\tau \tau}
\end{pmatrix}
\label{eq:params}
\end{equation}
with $\eta_{\alpha\beta}, \alpha_{\alpha \beta}\ll1$. Note that we choose to label the $\alpha$ matrix elements with flavour indices for notation ease instead of using numbers as in~\cite{Escrihuela:2015wra}. Furthermore, in~\cite{Escrihuela:2015wra} the identity matrix is not singled out from $\alpha$ as in our Eq.~\eqref{eq:params} so that the diagonal elements $\alpha_{ii}$ in~\cite{Escrihuela:2015wra} are close to $1$ instead of small. Therefore, in practice, $\alpha_{ii} \to 1 - \alpha_{\alpha \alpha}$. These changes are only cosmetic and the following discussion applies to~\cite{Escrihuela:2015wra} with the above-mentioned identification. 

The deviations from unitarity are directly related to the heavy-active neutrino mixing. For instance, in the hermitian parametrization one can directly identify~\cite{Antusch:2009pm}
\begin{equation} 
\eta = \frac{\Theta \Theta^\dagger}{2}
\end{equation}
where $\Theta = m_D^\dagger M^{-1}$ is the heavy-active mixing given by the ratio of the Dirac over the Majorana mass scales. Thus, $(1-\eta)$ is just the first term in the cosine series correcting the unitary rotation $U'$. It is also straightforward to obtain the relation between the heavy-active neutrino mixing and the $\alpha$ parameters in the triangular parametrization, if one considers that the heavy-active mixing can also be encoded by introducing additional complex rotations characterized by new mixing angles $\theta_{ij}$, with $j > 3$. For example,
\begin{equation}
U_{14} = \begin{pmatrix}
c_{14} & 0 & 0 & \hat{s}^*_{14} \\
0 & 1 & 0 & 0 \\
0 & 0 & 1 & 0 \\
-\hat{s}_{14} & 0 & 0 & c_{14}
\end{pmatrix},
\end{equation}
where $\hat{s}_{ij} = e^{i \delta_{ij}} s_{ij}$, $s_{ij} = \sin \theta_{ij}$ and $c_{ij} = \cos \theta_{ij}$. In the correct 
order, these extra rotations lead to a lower triangular matrix. For 3 extra 
neutrinos we can use $U_{36} U_{26} U_{16} U_{35} U_{25} U_{15} U_{34} U_{24} U_{14}$ (where we have not included unphysical rotations among the sterile neutrinos), leading to~\cite{Xing:2007zj}:
\begin{equation}
\alpha \simeq \begin{pmatrix}
\frac{1}{2} \left( s^2_{14} + s^2_{15} + s^2_{16} \right) & 0 & 0 \\
\hat{s}_{14}\hat{s}^*_{24} + \hat{s}_{15}\hat{s}^*_{25} + \hat{s}_{16}\hat{s}^*_{26} & \frac{1}{2} \left( s^2_{24} + s^2_{25} + s^2_{26} \right) & 0 \\
\hat{s}_{14}\hat{s}^*_{34} + \hat{s}_{15}\hat{s}^*_{35} + \hat{s}_{16}\hat{s}^*_{36} & \hat{s}_{24}\hat{s}^*_{34} + \hat{s}_{25}\hat{s}^*_{35} + \hat{s}_{26}\hat{s}^*_{36} & \frac{1}{2} \left( s^2_{34} + s^2_{35} + s^2_{36} \right) \\
\end{pmatrix}\, ,
\label{eq:sines}
\end{equation}
which is accurate to second order in the (small) extra mixing angles.

In principle, the two parametrizations should be equally valid. However, the alternative use of each of them seemingly leads 
to inconsistent results. As an illustrative example, let us compare the $\nu_\mu$ disappearance probability in the 
atmospheric regime in the two parametrizations, obtained at linear order in the non-unitarity parameters and for 
$\theta_{13} = \Delta_{21}=0$ 
\begin{eqnarray}
\mathcal P_{\mu\mu}^{\eta}&=&
1-\left\{\sin^2{2\theta'_{23}} -
2\text{Re}[\eta_{\mu\tau}]\sin{4 \theta'_{23}}\right\}\sin^2{\Delta_{31}}, 
\nonumber\\
\mathcal P_{\mu\mu}^{\alpha}&=&1-\sin^2{2\theta_{23}}\sin^2{\Delta_{31}}, 
\label{eq:Pmumu_alpha}
\end{eqnarray}
where $\Delta_{ij}=\Delta m^2_{ij}L/4E$. Here, $\mathcal P_{\alpha\beta}$ denotes the ``experimental'' oscillation probability in vacuum 
including the normalization factors as discussed in Sec.~\ref{sec:comparison}. 

The naive conclusion derived from Eq.~\eqref{eq:Pmumu_alpha} is that for the Hermitian parametrization good sensitivity to the non-unitarity parameter $\eta_{\mu\tau}$ is expected in this channel, since it appears at linear order. Conversely, the triangular parametrization does not show this effect. This apparent 
inconsistency stems from the fact that the unitary matrices $U$ and $U'$ are, in fact, different. This is the case even though these matrices are traditionally identified with the standard unitary PMNS matrix in each parametrization. However, this identification is only accurate up to the small corrections stemming from the deviations from unitarity. As we will show below, the differences between the two are indeed linear in the non-unitarity parameters, and the two matrices can be easily related to each other. The relevant question is therefore which of these matrices, if any, that more closely matches the one that is determined through the present neutrino oscillation data. Starting from Eq.~\eqref{eq:N} a unitary rotation $V$ can be performed to relate $U$ and $U'$
\begin{equation}
N = (I-\alpha) U = (I-\eta) V V^\dagger U' 
\end{equation}
and therefore 
\begin{equation}
I-\alpha = (I - \eta)V \quad \mathrm{and} \quad U = V^\dagger U'.
\label{eq:V}
\end{equation}
From the first relation in Eq.~\eqref{eq:V} the elements of $V$ can be identified as
\begin{equation}
V =
I - \begin{pmatrix}
0 & -\eta_{e \mu} & -\eta_{e \tau} \\
\eta^*_{e \mu} & 0 & -\eta_{\mu \tau} \\
\eta^*_{e \tau} & \eta^*_{\mu \tau} & 0
\end{pmatrix}
\end{equation}
at linear order in $\eta$. Substituting again in Eq.~\eqref{eq:V} the relations
\begin{equation}
\begin{pmatrix}
\alpha_{ee} & 0 & 0\\
\alpha_{\mu e} & \alpha_{\mu \mu} & 0\\
\alpha_{\tau e} & \alpha_{\tau \mu} & \alpha_{\tau \tau}
\end{pmatrix} =
\begin{pmatrix}
\eta_{ee} & 0 & 0 \\
2 \eta^*_{e \mu} & \eta_{\mu \mu} & 0 \\
2 \eta^*_{e \tau} & 2 \eta^*_{\mu \tau} &  \eta_{\tau \tau}
\end{pmatrix}
\label{eq:mapalpha}
\end{equation}
and
\begin{equation}
U = V^\dagger U'= \left( I + \begin{pmatrix}
0 & -\eta_{e \mu} & -\eta_{e \tau} \\
\eta^*_{e \mu} & 0 & -\eta_{\mu \tau} \\
\eta^*_{e \tau} & \eta^*_{\mu \tau} & 0
\end{pmatrix}
\right)
U' 
\label{eq:equality}
\end{equation}
are found.
This implies the following mapping between the two sets of mixing angles\footnote{Note that, apart from correcting the PMNS mixing angles and CP-phase $\delta_{\mathrm{CP}}$ at order $\eta$, 
phase redefinitions of the three charged leptons as well as corrections to the two neutrino Majorana phases are necessary
at the same order.} in $U'$ and $U$:
\begin{align}
\nonumber
\theta_{12}-\theta'_{12} &= \frac{\Rea(s_{23}\eta_{e \tau} - c_{23} \eta_{e \mu})}{c_{13}}, \\
\nonumber
\theta_{13}-\theta'_{13} &= \Rea(-s_{23} e^{i \delta_{\mathrm{CP}}} \eta_{e \mu} - c_{23} e^{i \delta_{\mathrm{CP}}} 
\eta_{e \tau}), \\
\nonumber
\theta_{23}-\theta'_{23} &= -\Rea(\eta_{\mu \tau}) + \tan{\theta_{13}} \Rea\left( c_{23} e^{i \delta_{\mathrm{CP}}} \eta_{e \mu}
- s_{23} e^{i \delta_{\mathrm{CP}}} \eta_{e \tau} \right), \\
\nonumber
 \delta_{\mathrm{CP}}-\delta'_{\mathrm{CP}}  &= \frac{\cos{2 \theta_{12}}}{s_{12}c_{12}c_{13}} \Ima\left(s_{23} \eta_{e\tau} - c_{23} 
\eta_{e \mu} \right) + \frac{1}{s_{13}c_{13}} \Ima\left(s_{23} e^{i \delta_{\mathrm{CP}}} \eta_{e \mu} + c_{23} 
e^{i \delta_{\mathrm{CP}}} \eta_{e \tau} \right) \\
&\phantom = -\frac{\tan\theta_{13}}{s_{23}c_{23}} \Ima\left(c^3_{23} e^{i \delta_{\mathrm{CP}}} \eta_{e \mu} + s^3_{23} 
e^{i \delta_{\mathrm{CP}}} \eta_{e \tau} + \frac{\eta_{\mu \tau}}{\tan\theta_{13}}\right).  
\label{eq:mapth}
\end{align}
When the relations given in Eqs.~\eqref{eq:mapalpha} and \eqref{eq:mapth} are taken into account the predictions for the 
different oscillation channels coincide at leading order in the non-unitarity parameters, as they should. 
An important conclusion derived from this is that the determination of the mixing angles themselves will generally be affected by non-unitarity corrections. However, the size of these corrections is, at present, negligible compared to the current uncertainties on the determination of the mixing angles themselves. These corrections are parametrization-dependent but, when taken into account and propagated consistently, the predictions derived from both schemes agree. 

For neutrino oscillation studies it seems advantageous to adopt the triangular parametrization, since it leads to fewer 
corrections given its structure. For instance, in the example shown in Eq.~\eqref{eq:Pmumu_alpha}
there are no corrections coming from non-unitarity for this parametrization, and thus the angle 
$\theta_{23}$ in $U$ can be identified with the angle determined in present global fits to a good approximation. 
Indeed, this is also the case for $\theta_{12}$ and $\theta_{13}$, since the $\mathcal P_{ee}$ oscillation probabilities 
in the solar regime (KamLAND) and in the atmospheric regime (Daya Bay, RENO, Double-Chooz) are also
independent of any non-unitarity corrections at linear order when the triangular parametrization is considered 
and when the appropriate normalization is taken into account, see Sec.~\ref{sec:comparison}. 

Thus, the $U$ matrix from the triangular parametrization corresponds, to a good approximation, with the unitary matrix 
obtained when determining $\theta_{12}$, $\theta_{23}$ and $\theta_{13}$ through present (disappearance) neutrino oscillation measurements. Since we are here interested in the impact of non-unitarity and sterile neutrinos on neutrino oscillation phenomenology 
we will therefore use the triangular parametrization in the remainder of this work.

As we will see in Sec.~\ref{sec:simulations}, the dependence on the diagonal non-unitarity parameters $\alpha_{\beta\beta}$ 
is particularly interesting. Indeed, when the normalization accounting for the new physics effects at the near detector is considered, it 
effectively cancels any leading order dependence on $\alpha_{\beta\beta}$ in disappearance channels in vacuum. This 
can be easily understood by introducing the triangular parameterization in Eq.~(\ref{eq:experimentalprobability}). Expanding in $\alpha_{\delta\gamma}$ we obtain
\be
\mathcal P_{\alpha\beta}=\left|\left(1+\alpha_{\alpha\alpha}-\alpha_{\beta\beta}\right)(U S^0 U^\dagger)_{\alpha\beta}  
-\sum_{\delta\neq\alpha}\alpha_{\alpha\delta}(U S^0 U^\dagger)_{\delta\beta}-
\sum_{\delta\neq\beta}(U S^0 U^\dagger)_{\alpha\delta}\alpha^*_{\beta\delta}
+\mathcal{O}\left(\alpha^2\right)\right|^2.
\label{eq:cancellation}
\ee
Therefore, when $\alpha=\beta$ the dependence on $\alpha_{\beta\beta}$ cancels out. This illustrates how relevant the role of 
the near detectors is regarding the sensitivity to the new physics parameters.

\section{NSI}
\label{sec:NSI}

Both types of new physics effects in neutrino oscillations discussed above can be described through the Non-Standard 
Interaction (NSI) formalism, which parametrizes the new physics effects
in neutrino production, detection, and propagation processes in a completely model-independent way. Let us first focus on NSI affecting neutrino production and detection. When these effects are included, the oscillation probability is given by 
\begin{equation}
P_{\alpha\beta} = |\left[(1+\epsilon^{d})U S^0 U^\dagger(1+\epsilon^{s})\right]_{\beta\alpha}|^2,
\label{eq:NSIoscillationformula}
\end{equation}
where $\epsilon^{s}$ and $\epsilon^{d}$ are general $3\times3$ complex matrices which represent the NSI modifications 
to the production and detection diagrams, respectively. $S^{0}$ is defined in Eq.~\eqref{S0} with the Hamiltonian $H$
given in Eq.~(\ref{eq:Hnonunitarity}). The non-unitarity (Eq.~\eqref{eq:nonunitarityoscillationformula}) and 
averaged-out sterile neutrino (Eq.~\eqref{eq:sterileoscillationformula}) effects at production and detection can be 
mapped to the NSI formalism (Eq.~\eqref{eq:NSIoscillationformula}) with the identification
\be
\epsilon_{\beta\alpha}^{s*}=\epsilon^{d}_{\alpha\beta}=-\alpha_{\alpha\beta}.
\ee
This mapping can be easily obtained just considering the triangular parameterization, which can be applied in both
the non-unitarity and averaged-out sterile neutrino cases, in Eqs.~\eqref{eq:nonunitarityoscillationformula} or~\eqref{eq:sterileoscillationformula} and comparing the result to Eq.~\eqref{eq:NSIoscillationformula}.

On the other hand, NSI affecting neutrino propagation are usually described through 
the Hamiltonian
\begin{equation}
\label{eq:HNSI}
H = \frac{1}{2E}\begin{pmatrix}
0 & 0 & 0 \\
0 & \Delta m_{21}^2 & 0 \\
0 & 0 & \Delta m_{31}^2
\end{pmatrix} +  V_{\rm CC}\, U^\dagger\begin{pmatrix}
1+\epsilon_{ee} & \epsilon_{e\mu} & \epsilon_{e\tau} \\
\epsilon^*_{e \mu} & \epsilon_{\mu\mu} & \epsilon_{\mu\tau} \\
\epsilon^*_{e \tau} & \epsilon^*_{\mu \tau} & \epsilon_{\tau\tau}
\end{pmatrix} U,
\end{equation}
in the mass basis, where $U$ is the standard unitary PMNS matrix, and $\epsilon_{\alpha\beta}$ and $\epsilon_{\alpha\alpha}$ are 
complex and real parameters respectively. In order to understand how the non-unitarity/averaged-out sterile neutrino corrected matter effects can be translated to this parametrization, we introduce the triangular parameterization of $N$ into Eq.~\eqref{eq:Hnonunitarity}, obtaining the following Hamiltonian
at leading order in $\alpha$
\begin{equation}
\label{eq:Hnonunitarityearth}
H = \frac{1}{2E}\begin{pmatrix}
0 & 0 & 0 \\
0 & \Delta m_{21}^2 & 0 \\
0 & 0 & \Delta m_{31}^2
\end{pmatrix} + \frac{V_{\rm CC}}{2}\,U^\dagger  \begin{pmatrix}
2-2\alpha_{ee}  & \alpha^*_{\mu e} & \alpha^*_{\tau e} \\
\alpha_{\mu e} & 2\alpha_{\mu\mu} & \alpha^*_{\tau \mu} \\
\alpha_{\tau e} & \alpha_{\tau \mu} & 2\alpha_{\tau\tau}
\end{pmatrix} U,
\end{equation}
where approximately equal densities of electrons $n_e$ and neutrons $n_n$ (for the neutral current contribution) in the Earth have been assumed (see also Ref.~\cite{Antusch:2008tz}).  Comparing Eqs.~\eqref{eq:Hnonunitarityearth} and~\eqref{eq:HNSI} we find the mapping between the NSI parametrization and the lower triangular parametrization of the non-unitarity and sterile neutrino scenarios
\begin{align}
\epsilon_{ee}&=-\alpha_{ee},  &  \epsilon_{\mu\mu}&=\alpha_{\mu\mu},  &
\epsilon_{\tau\tau}&=\alpha_{\tau\tau}, \nonumber \\
\epsilon_{e\mu}&=\frac 12 \alpha^*_{\mu e}, & \epsilon_{e\tau}&=\frac 12 \alpha^*_{\tau e}, & \epsilon_{\mu \tau}&=\frac 12 \alpha^*_{\tau \mu},  
\end{align}
which apply for neutrino oscillation experiments in the Earth with constant matter. Note that, in presence of production and detection NSI, the same normalization as for the non-unitarity case discussed in Section~\ref{sec:comparison} needs to be taken into account.  

\section{Present constraints on deviations from unitarity}
\label{sec:bounds}

The mapping to NSI described above works both for the non-unitarity and the averaged out sterile neutrino contributions 
to neutrino oscillations. However, the present constraints on each of these contributions from other observables are 
very different. Indeed, PMNS non-unitarity from very heavy extra neutrinos induces modifications of the $W$ and $Z$ 
couplings that impact precision electroweak and flavour observables~\cite{Shrock:1980vy,Schechter:1980gr,Shrock:1980ct,Shrock:1981wq,Langacker:1988ur,Bilenky:1992wv,Nardi:1994iv,Tommasini:1995ii,Antusch:2006vwa,Antusch:2008tz,Biggio:2008in,Alonso:2012ji,Antusch:2014woa,Abada:2015trh,Abada:2016awd,Fernandez-Martinez:2016lgt}. These modification translate into very strong upper limits on the $\alpha$ parameters. 
These have been taken from Ref.~\cite{Fernandez-Martinez:2016lgt} and are listed in the left column in Table~\ref{tab:bounds}. The second number quoted in parenthesis for the $\alpha_{\mu e}$ element includes the $\mu \to e \gamma$ observable, which can in principle be evaded~\cite{Forero:2011pc} for heavy neutrino masses close to $M_W$ and some fine-tuning of the parameters. In this case, the quoted bound is derived from the constraints on the diagonal parameters, through Eq.~\eqref{eq:sines}.
%
\begin{table}[t!]
\setlength{\tabcolsep}{7pt}
\begin{center}
\renewcommand{\arraystretch}{1.6}
\begin{tabular}{|  c@{\quad} | c@{\quad} | c@{\quad} c@{\quad}   | }
\hline
    & ``Non-Unitarity'' & \multicolumn{2}{c|}{``Light steriles''} \\
     & ($m>$ EW)  &  $\Delta m^2 \gtrsim 100$~eV$^2$ & $\Delta m^2 \sim 0.1-1$~eV$^2$\\ \hline\hline
$\alpha_{ee} $ & $1.3 \cdot 10^{-3}$~\cite{Fernandez-Martinez:2016lgt} & $2.4 \cdot 10^{-2} $~\cite{Declais:1994su} & $1.0 \cdot 10^{-2} $~\cite{Declais:1994su}\\
$\alpha_{\mu\mu}$ & $2.2 \cdot 10^{-4}$~\cite{Fernandez-Martinez:2016lgt} & $2.2 \cdot 10^{-2}$~\cite{Abe:2014gda} & $1.4 \cdot 10^{-2}$~\cite{MINOS:2016viw}\\
$\alpha_{\tau\tau}$ & $2.8 \cdot 10^{-3}$~\cite{Fernandez-Martinez:2016lgt} & $1.0 \cdot 10^{-1}$~\cite{Abe:2014gda} & $1.0 \cdot 10^{-1}$~\cite{Abe:2014gda}\\
$\lvert\alpha_{\mu e}\rvert$ & $6.8 \cdot 10^{-4} \; (2.4 \cdot 10^{-5})$~\cite{Fernandez-Martinez:2016lgt} & $2.5 \cdot 10^{-2} $~\cite{Astier:2003gs} & $1.7 \cdot 10^{-2} $ \\
$\lvert\alpha_{\tau e}\rvert$ & $2.7 \cdot 10^{-3}$~\cite{Fernandez-Martinez:2016lgt} & $6.9 \cdot 10^{-2}$ & $4.5 \cdot 10^{-2}$ \\
$\lvert\alpha_{\tau\mu}\rvert$ & $1.2 \cdot 10^{-3}$~\cite{Fernandez-Martinez:2016lgt} & $1.2 \cdot 10^{-2}$~\cite{Astier:2001yj} & $5.3 \cdot 10^{-2}$ \\ \hline\hline
\end{tabular}
\caption{\label{tab:bounds} Current upper bounds on the $\alpha$ parameters in the scenarios considered in this work. The limits are shown at $2\sigma$ and 95\% CL (1 d.o.f.) for the non-unitarity and light sterile neutrino
scenarios. The bounds in the middle column apply for $\Delta m^2 \gtrsim 100$~eV$^2$ and will thus be relevant when the sterile neutrino oscillations are in the averaged-out regimes for both the near and far detectors of most long-baseline experiments. The bounds in the right column apply for $\Delta m^2 \sim 0.1-1$~eV$^2$ and will thus be relevant when the sterile neutrino oscillations are in the averaged-out regime for the far detector, but not for the near detector. The second number quoted in parenthesis for the $\alpha_{\mu e}$ element includes the $\mu \to e \gamma$ observable, which can in principle be evaded~\cite{Forero:2011pc}
, see main text for details. The numbers for the off-diagonal parameters without a reference are obtained indirectly from 
constraints on the diagonal parameters via $\alpha_{\alpha \beta} \leq 2 \sqrt{\alpha_{\alpha \alpha} \alpha_{\beta \beta}}$
(see Eq.~\eqref{eq:sines}). See Appendix~\ref{app} for further details.}
\end{center}
\end{table}

However, for sterile neutrinos with masses below the electroweak scale these stringent constraints are lost, since all mass eigenstates are kinematically available in the observables used to derive the constraints and unitarity is therefore restored. If the masses of the extra states are in the MeV or GeV range, even stronger constraints can be derived from direct searches at beam-dump experiments as well as from meson and beta decays~\cite{Atre:2009rg,Ruchayskiy:2011aa,Drewes:2015iva}. On the other hand, for masses below the keV scale even the beta decay searches are no longer sensitive, and the only applicable bounds are the much milder constraints stemming from the non-observation of their effects in neutrino oscillation experiments~\cite{Kopp:2013vaa,Parke:2015goa,Dutta:2016glq}. The sensitivity, or lack thereof, of oscillation experiments to sterile neutrino mixing will depend on the actual value of the sterile neutrino mass, which determines if the corresponding $\Delta m^2$ leads to oscillations for the energy and baseline that characterize the experimental setup. As $\Delta m^2$ increases, there will be a point at which the sterile neutrino oscillations enter the averaged-out regime. Once oscillations are averaged-out, the constraints derived will become independent of $\Delta m^2$ and apply to arbitrarily large values of $\Delta m^2$. Limits derived in this regime are summarized in in the middle column of Table~\ref{tab:bounds} and apply as long as $\Delta m^2 >100$~eV$^2$. They are thus relevant when the sterile neutrino oscillations are in the averaged out regime for both the near and far detectors of the DUNE experiment. Some of these constraints also apply for values of $\Delta m^2$ smaller than 100~eV$^2$. For a more comprehensive breakdown of the available constraints and their ranges of applicability, we refer the interested reader to Appendix~\ref{app}.

Even though the case in which the sterile neutrino oscillations are undeveloped at the near detector, but 
averaged-out at the far, applies to a significantly smaller fraction of the parameter space, we find it instructive 
to analyze this scenario as well, since it leads to very different phenomenology and sensitivities, as will be 
discussed in Section~\ref{sec:simulations}. For the case of DUNE, this scenario requires $\Delta m^2 \sim 0.1-1$~eV$^2$ and 
the corresponding constraints have been compiled in the right column of Table~\ref{tab:bounds}. Notice that in this range of $\Delta m^2$ most constraints come from experiments that would not have reached the averaged-out regime but would rather have oscillations well-matched to their energies and baselines. Thus, the corresponding constraints in this regime oscillate significantly and the value quoted in the table is the most conservative available in that range.

\section{DUNE sensitivities}
\label{sec:simulations}

In this section we present, as an example, the sensitivities that the proposed DUNE experiment would have to PMNS 
non-unitarity or, equivalently, to averaged-out sterile neutrino oscillations as discussed in 
Section~\ref{sec:comparison}. For this analysis we choose the triangular parametrization of the new physics effects 
since, as argued in Section~\ref{sec:param}, its unitary part can be more directly mapped to the ``standard'' 
PMNS matrix as determined by present experiments through neutrino oscillation disappearance channels. Indeed, 
production and detection new physics effects in a given channel $P_{\alpha \beta}$ only depend on the elements 
$\alpha_{\gamma \rho}$ such that $\gamma, \rho \leq \alpha$ or $\gamma, \rho \leq \beta$ when the flavour indices 
are ordered as $e < \mu < \tau$~\cite{Escrihuela:2015wra}. Furthermore, when the new physics affects the near and 
far detectors in the same manner, the normalization of the oscillation probabilities presented in 
Eq.~\eqref{eq:experimentalprobability} has to be applied, which effectively cancels any leading order dependence 
on the new physics parameters $\alpha_{\beta\beta}$ in disappearance channels in vacuum (see Eq.~(\ref{eq:cancellation})). The choice of the facility under 
study is motivated by the strong matter effects that characterize the DUNE setup and that allow to probe not 
only the source and detector effects induced by the new physics, but also the matter effects which now provide 
sensitivity to other $\alpha_{\gamma \rho}$ parameters not necessarily satisfying  
$\gamma, \rho \leq \alpha$ or $\gamma, \rho \leq \beta$.

The simulation of the DUNE setup was performed with the GLoBES software~\cite{Huber:2004ka,Huber:2007ji} using the DUNE CDR configuration presented in Ref.~\cite{Alion:2016uaj}. The new physics effects have been implemented in GLoBES via the MonteCUBES~\cite{Blennow:2009pk}\footnote{A new version of the MonteCUBES software implementing the triangular parametrization is available.} plug-in, which has also been used to perform a Markov chain Monte Carlo (MCMC) scan of the 15-dimensional parameter space (the 6 standard oscillation parameters plus the 6 moduli of the $\alpha_{\gamma \rho}$ non-zero entries and the 3 phases of the off-diagonal elements). In the fit, the assumed true values for the standard oscillation parameters are set according to their current best-fits from Ref.~\cite{Gonzalez-Garcia:2014bfa}. The mixing angles and squared-mass splittings are allowed to vary in the simulations, using a Gaussian prior corresponding to their current experimental uncertainties from Ref.~\cite{Gonzalez-Garcia:2014bfa} centered around their true values. In the case of $\theta_{13}$ and $\theta_{23}$ the Gaussian priors are implemented on $\sin^2 2 \theta$, which is a more accurate description of the present situation and, in the case of $\theta_{23}$, allows to properly account for the octant degeneracy: $\Delta m^2_{21}=(7.50\pm 0.18)\cdot 10^{-5}$~eV$^2$, $\Delta m^2_{31}=(2.457\pm 0.049)\cdot 10^{-3}$~eV$^2$, $\theta_{12}= 33.48^\circ \pm 0.77^\circ$, $\sin^22\theta_{13}=0.085\pm 0.005$, $\sin^22\theta_{23}=0.991\pm 0.02$. Notice that, as described in Section~\ref{sec:param}, the use of the triangular parametrization allows a direct mapping of the present measurements to the elements of the $U$ matrix. Nevertheless, the present uncertainties adopted in this analysis are still large enough that any correction due to non-unitarity is negligible. The CP-violating phase is left completely free during the simulations, and its true value is set to $\delta_\mathrm{CP}=-\pi/2$. Finally, a 2~\% uncertainty in the PREM matter density profile~\cite{Dziewonski:1981xy} has also been considered. 

We have performed simulations for two distinct new physics scenarios. In the first case (ND averaged) we normalize the oscillation probabilities according to Eq.~\eqref{eq:experimentalprobability}. Indeed, as discussed in Section~\ref{sec:comparison}, at leading order in the new physics parameters this scenario accurately describes both the effects of PMNS non-unitarity from very heavy neutrinos as well as sterile neutrino oscillations that have been averaged out both at the near detector (ND) and far detector. For the DUNE setup, the requirement for having averaged-out oscillations at the near detector translates to the condition $\Delta m^2 >$~few 100 eV$^2$. The second scenario ({ND undeveloped}) would correspond to the case where sterile neutrino oscillations are averaged out at the far detector but have not yet developed at the near detector. In this case, no extra normalization is needed and the oscillation probability is directly given by Eq.~\eqref{eq:nonunitarityoscillationformula}. Note that, for the energies and baseline characterizing the DUNE setup, only values of the sterile neutrino masses around $\Delta m^2 \sim 0.1-1$~eV$^2$ roughly satisfy these conditions. However, we find it instructive to study also this regime in order to remark the differences between the two scenarios and the importance of the normalization in Eq.~\eqref{eq:experimentalprobability} that will generally apply in most of the parameter space.
\begin{figure}
 \includegraphics[width=0.48\columnwidth]{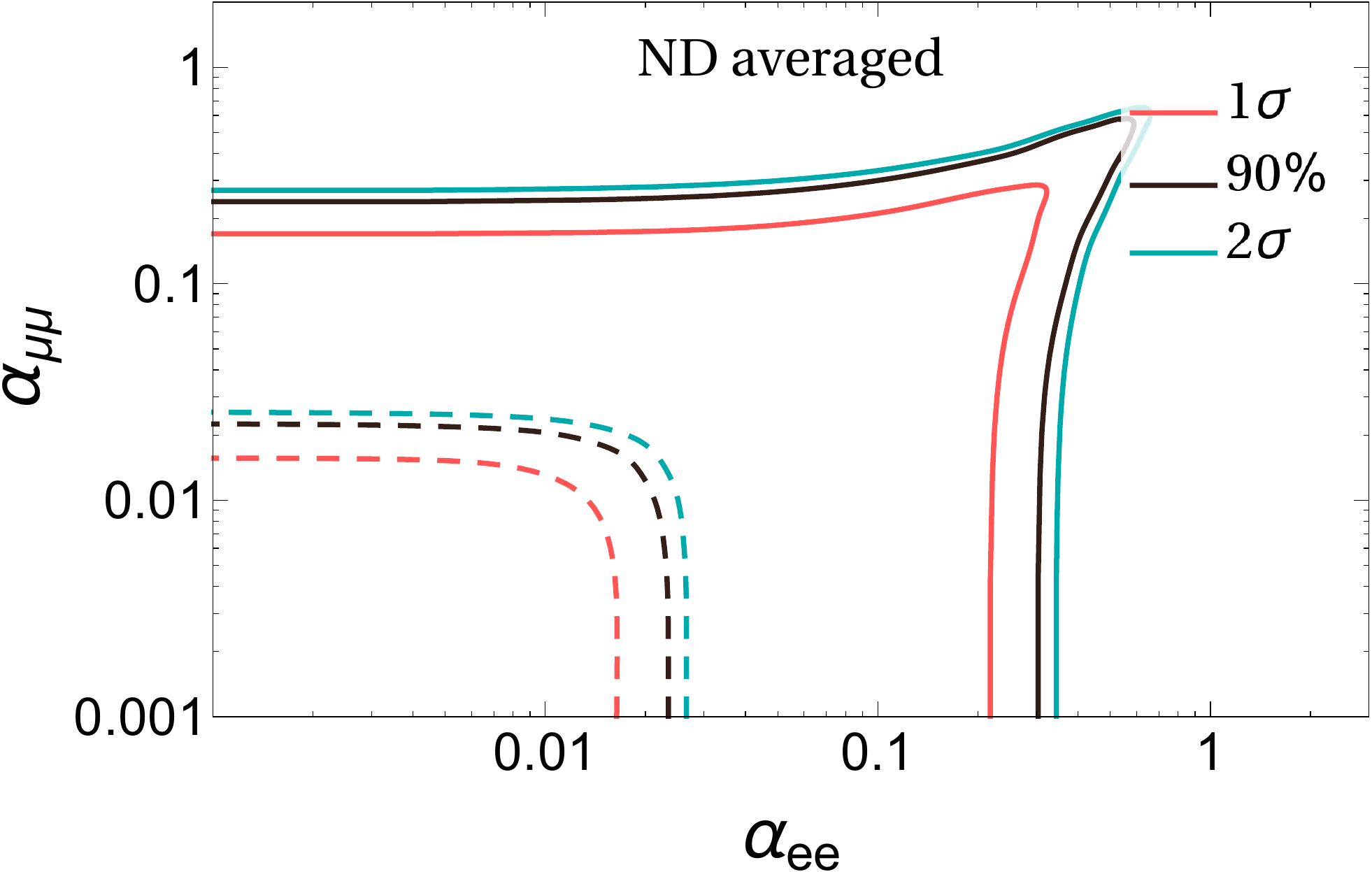} 
 \includegraphics[width=0.48\columnwidth]{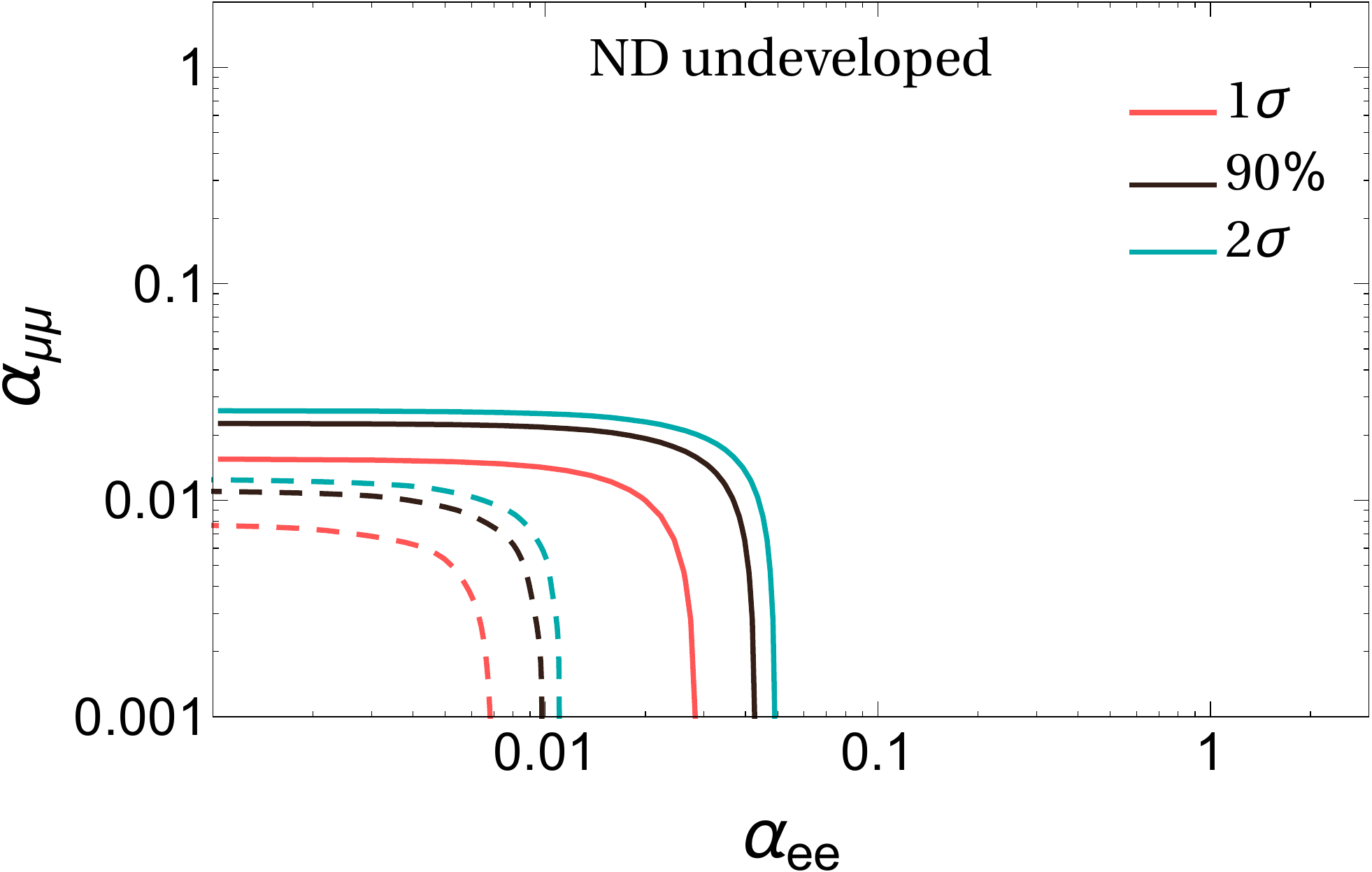} \\
 \includegraphics[width=0.48\columnwidth]{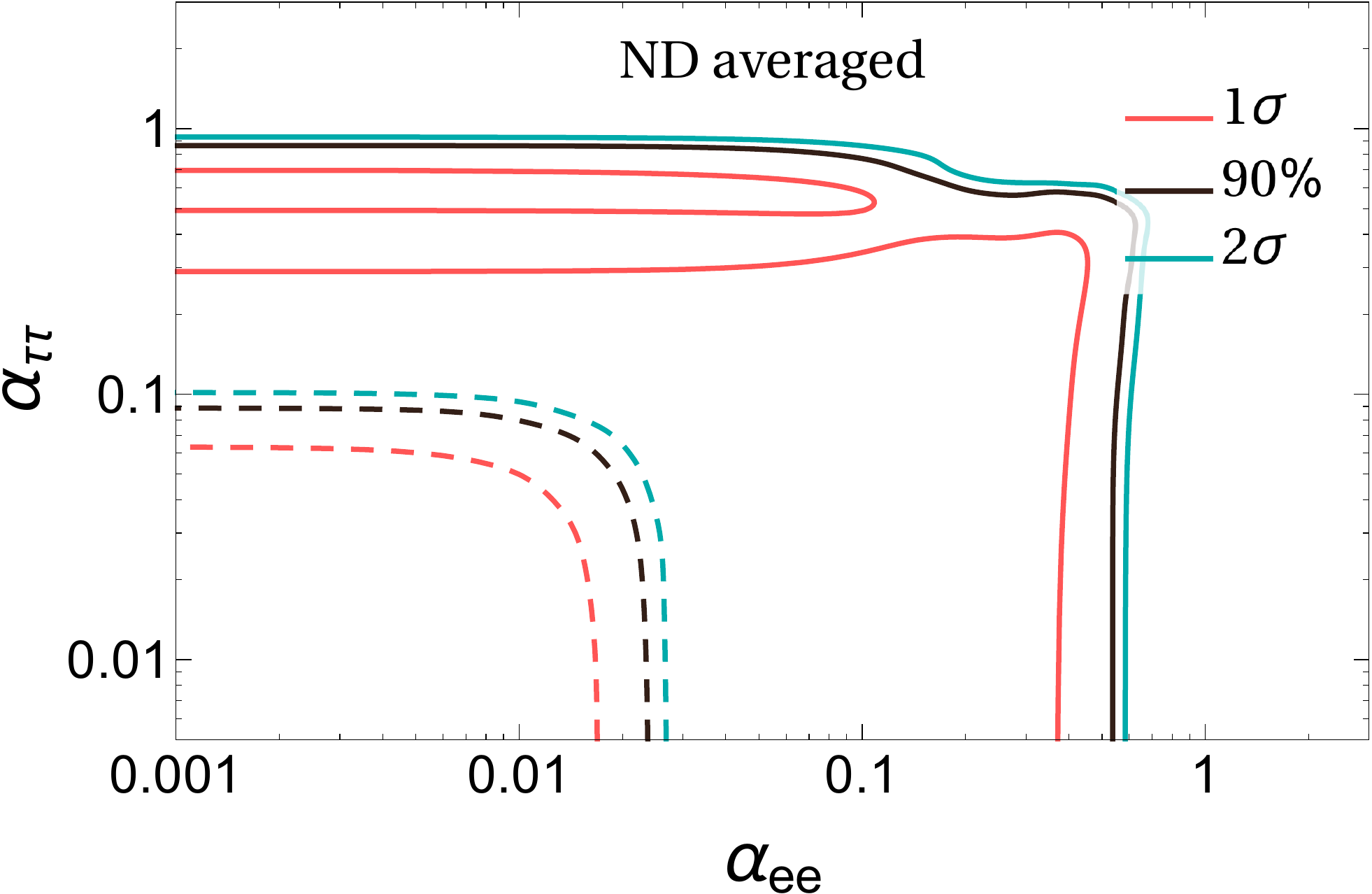} 
 \includegraphics[width=0.48\columnwidth]{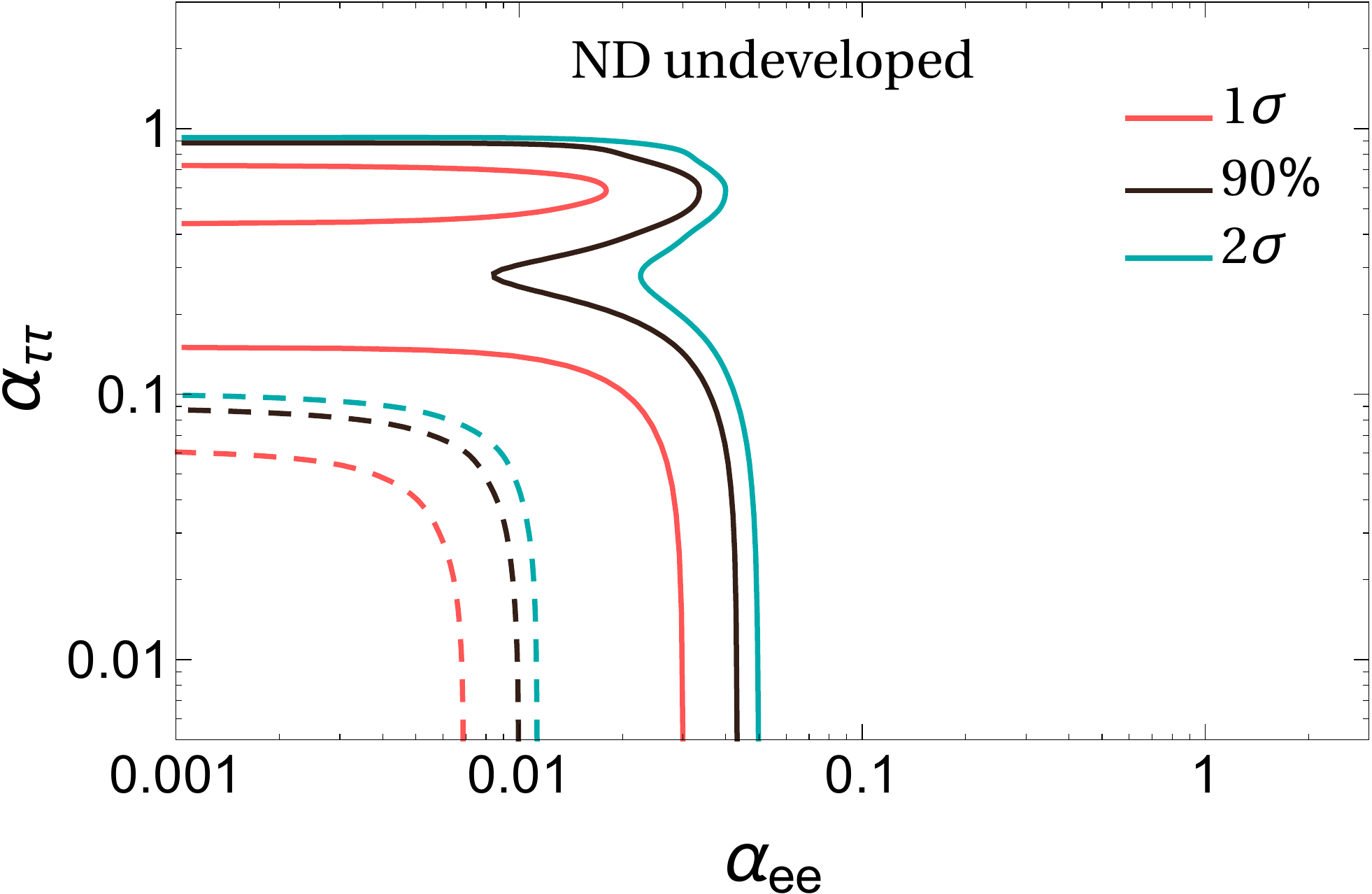} \\
  \includegraphics[width=0.48\columnwidth]{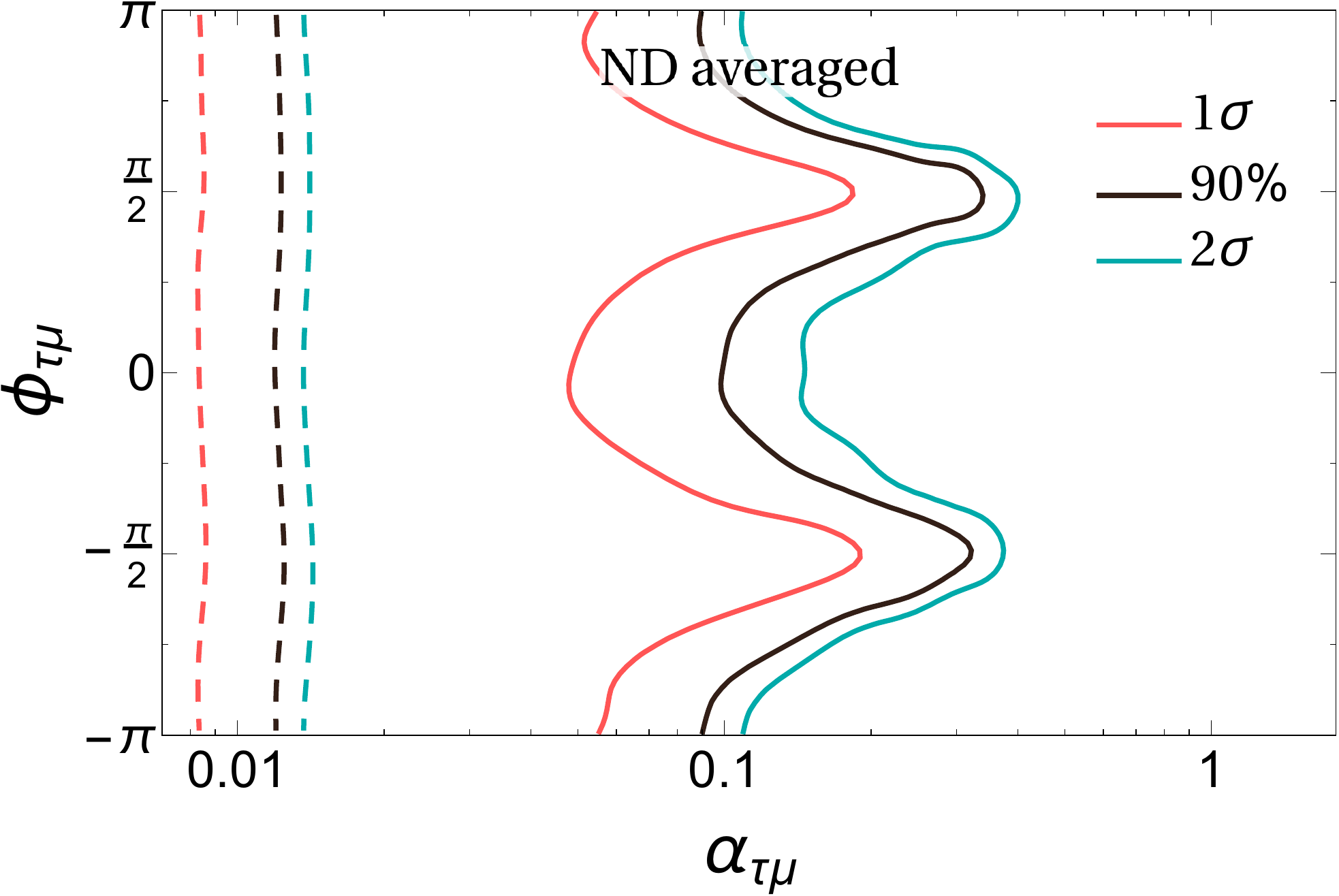} 
 \includegraphics[width=0.48\columnwidth]{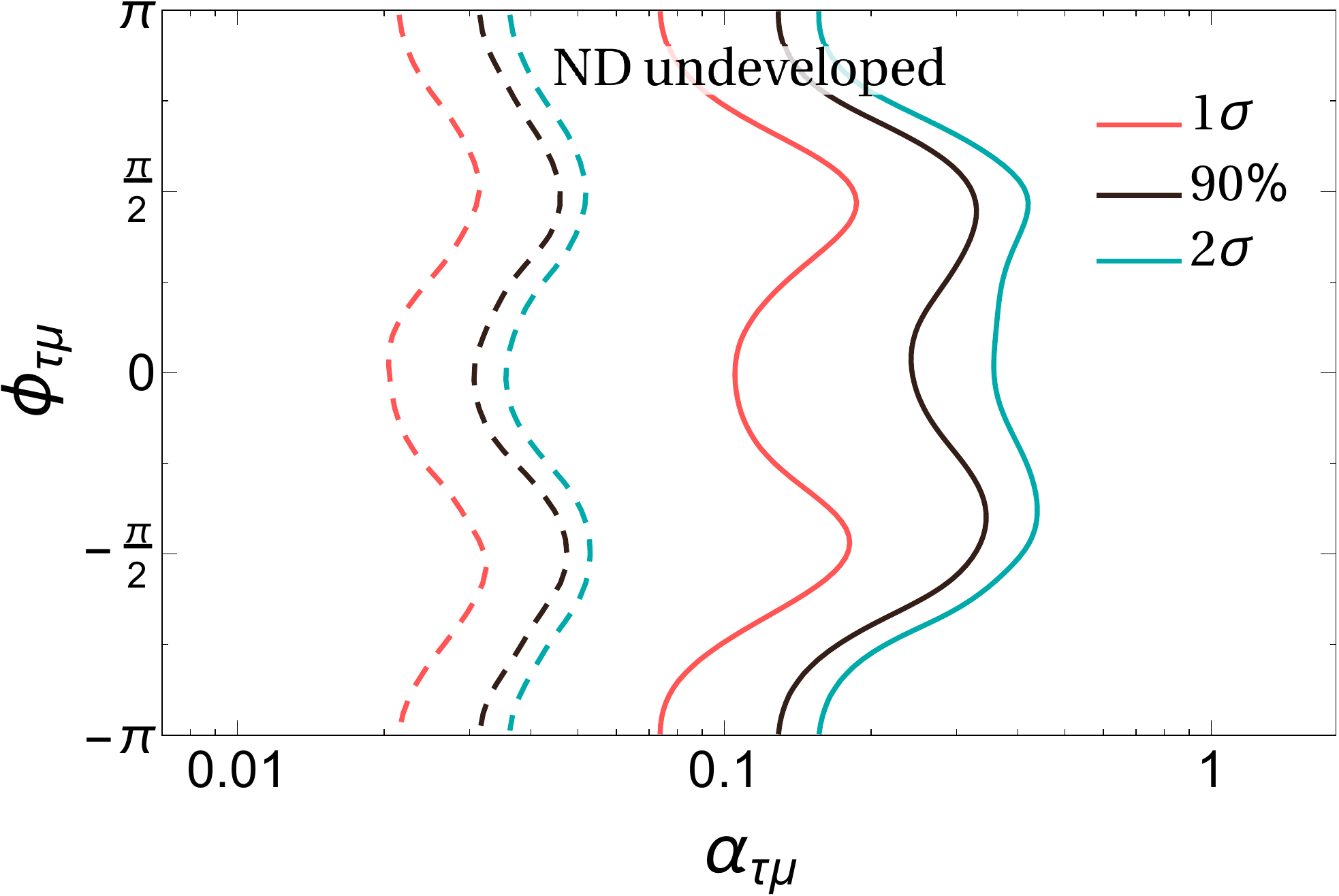}
 \begin{center}
  \caption{Expected frequentist allowed regions at the $1 \sigma$, $90\%$ and $2 \sigma$ C.L.\ for DUNE. All new physics parameters are assumed to be zero so as to obtain the expected sensitivities. The left panels (ND averaged) correspond to the non-unitarity case, or to the sterile case when the light-heavy oscillations are averaged out in the near and far detectors. The right panels (ND undeveloped) give the sensitivity for the sterile case when the light-heavy oscillations have not yet developed in the near detector, but are averaged out in the far. The solid lines correspond to the analysis of DUNE data alone, while the dashed lines include the present constraints on sterile neutrino mixing from the middle and right columns in Tab.~\ref{tab:bounds} for the NS averaged and ND undeveloped scenarios respectively. \label{fig:plots}}
 \end{center}
\end{figure}

\begin{figure}
 \includegraphics[width=0.48\columnwidth]{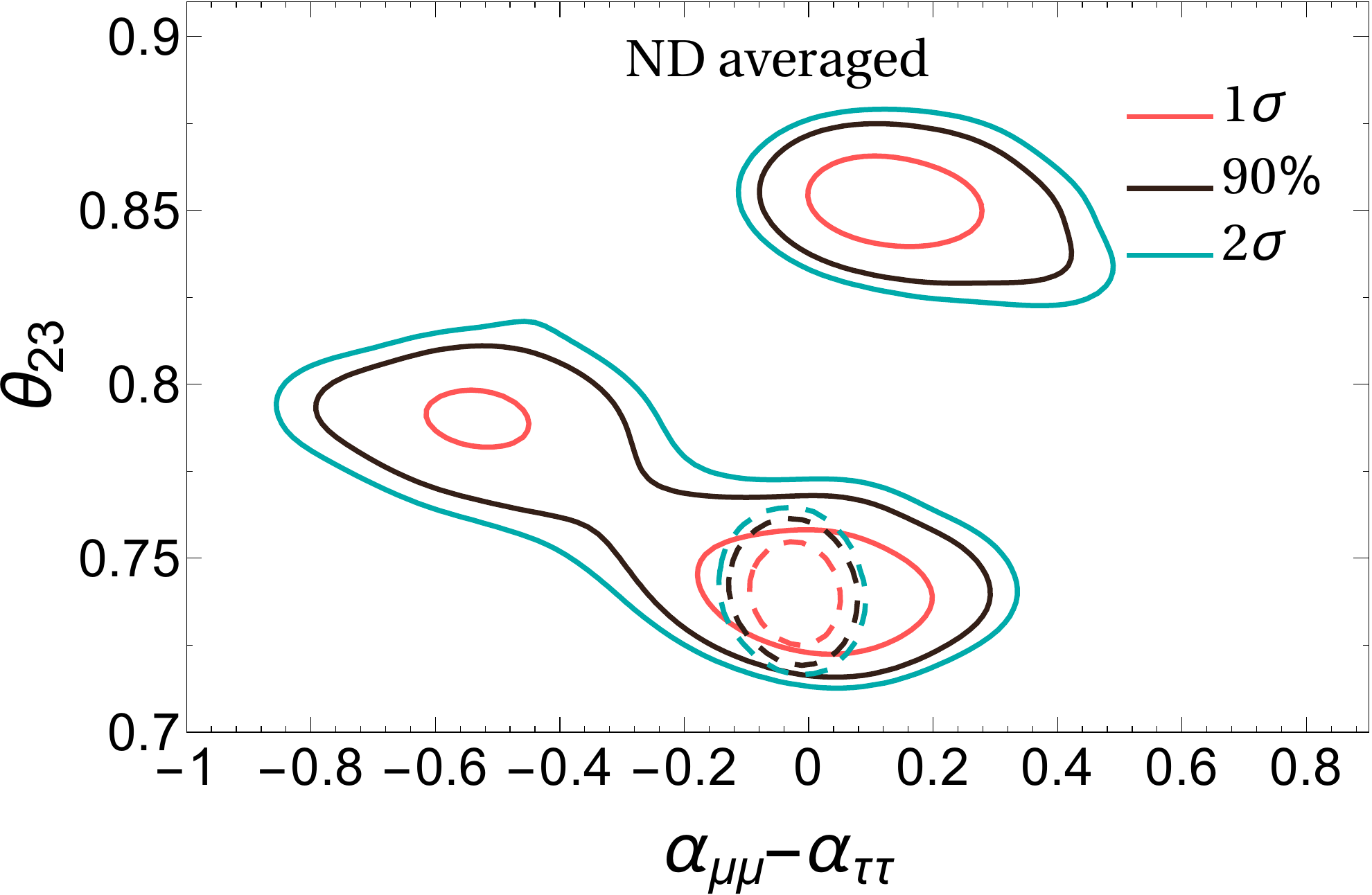} 
 \includegraphics[width=0.48\columnwidth]{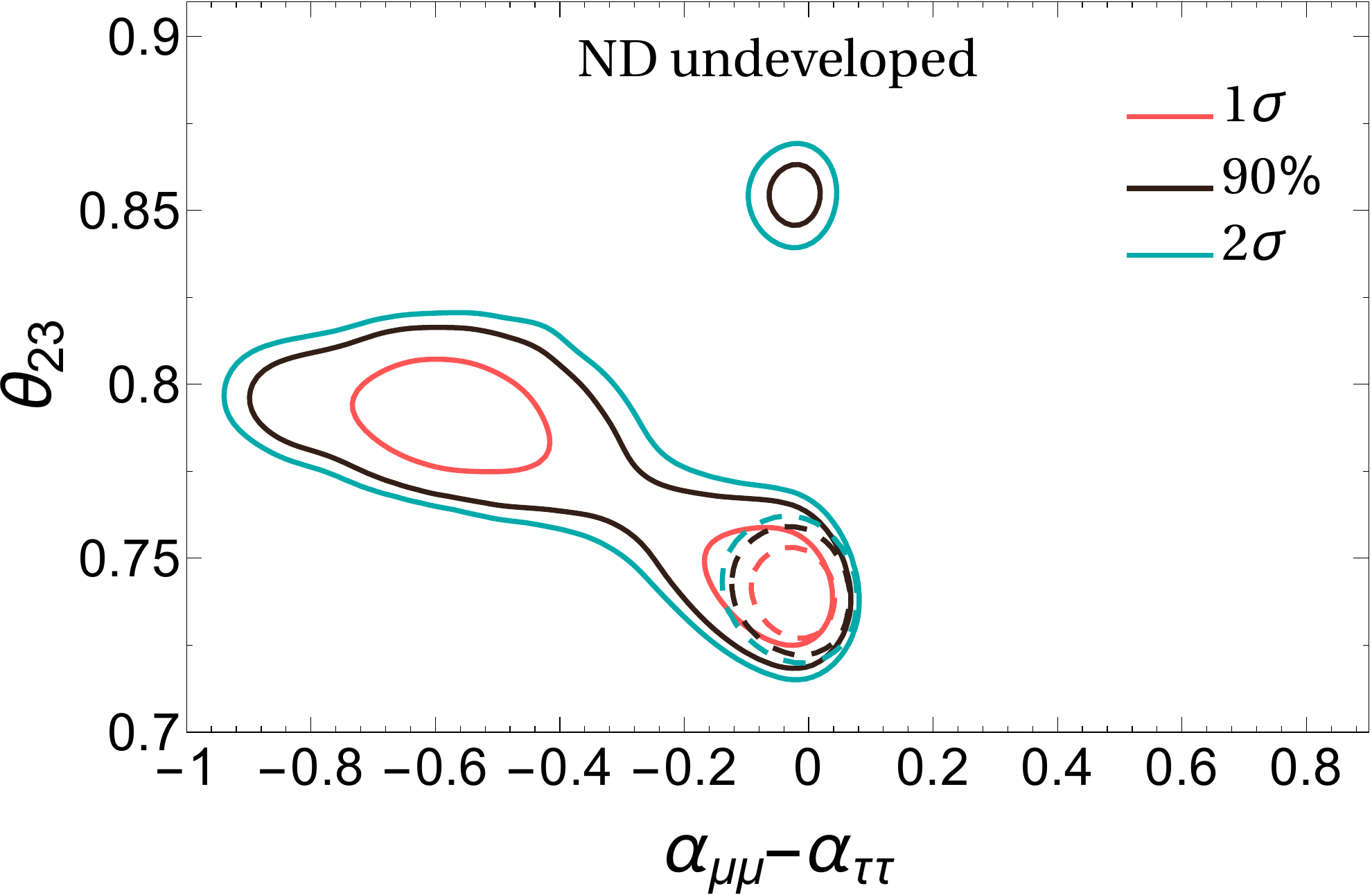} \\
 \includegraphics[width=0.48\columnwidth]{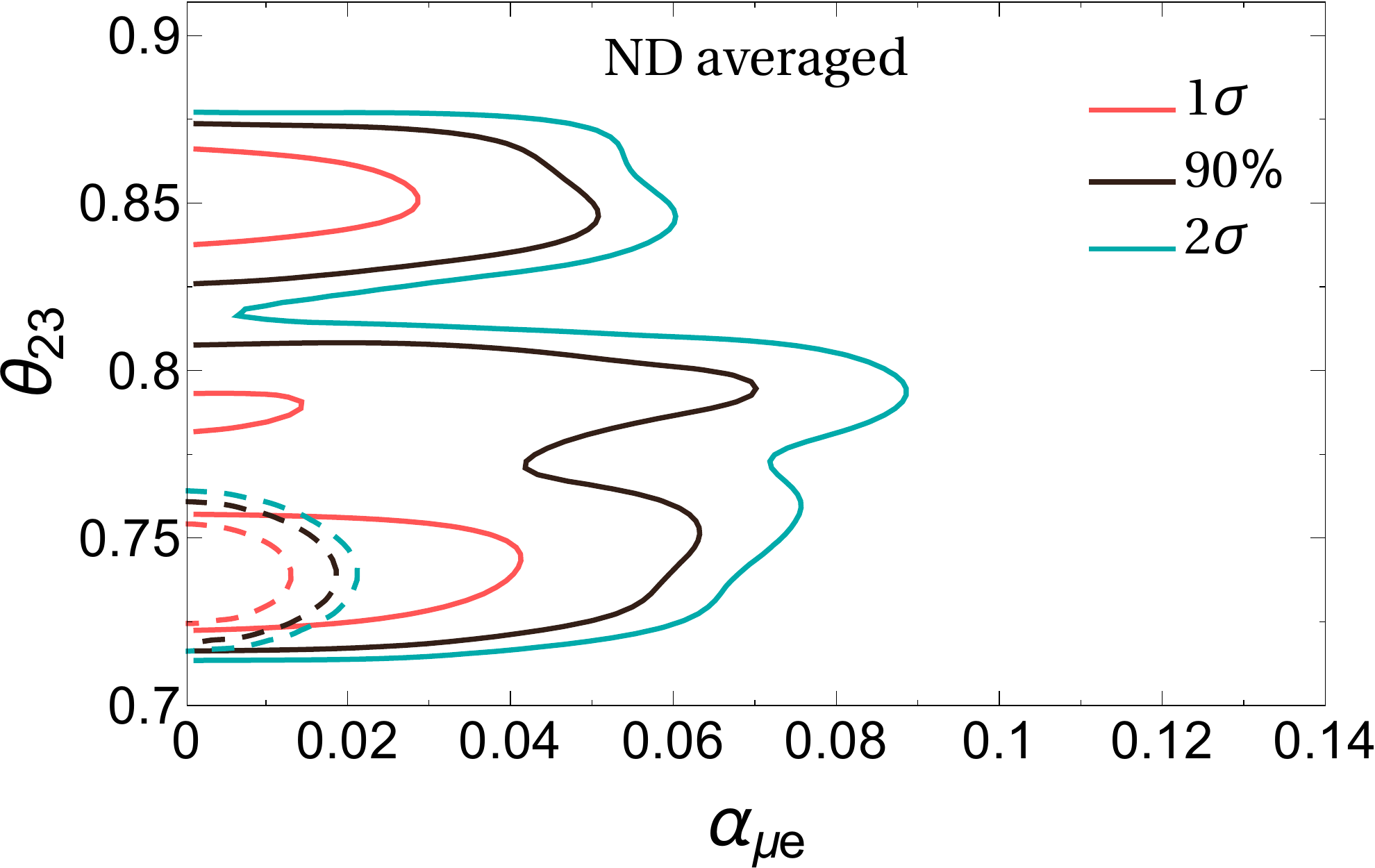} 
 \includegraphics[width=0.48\columnwidth]{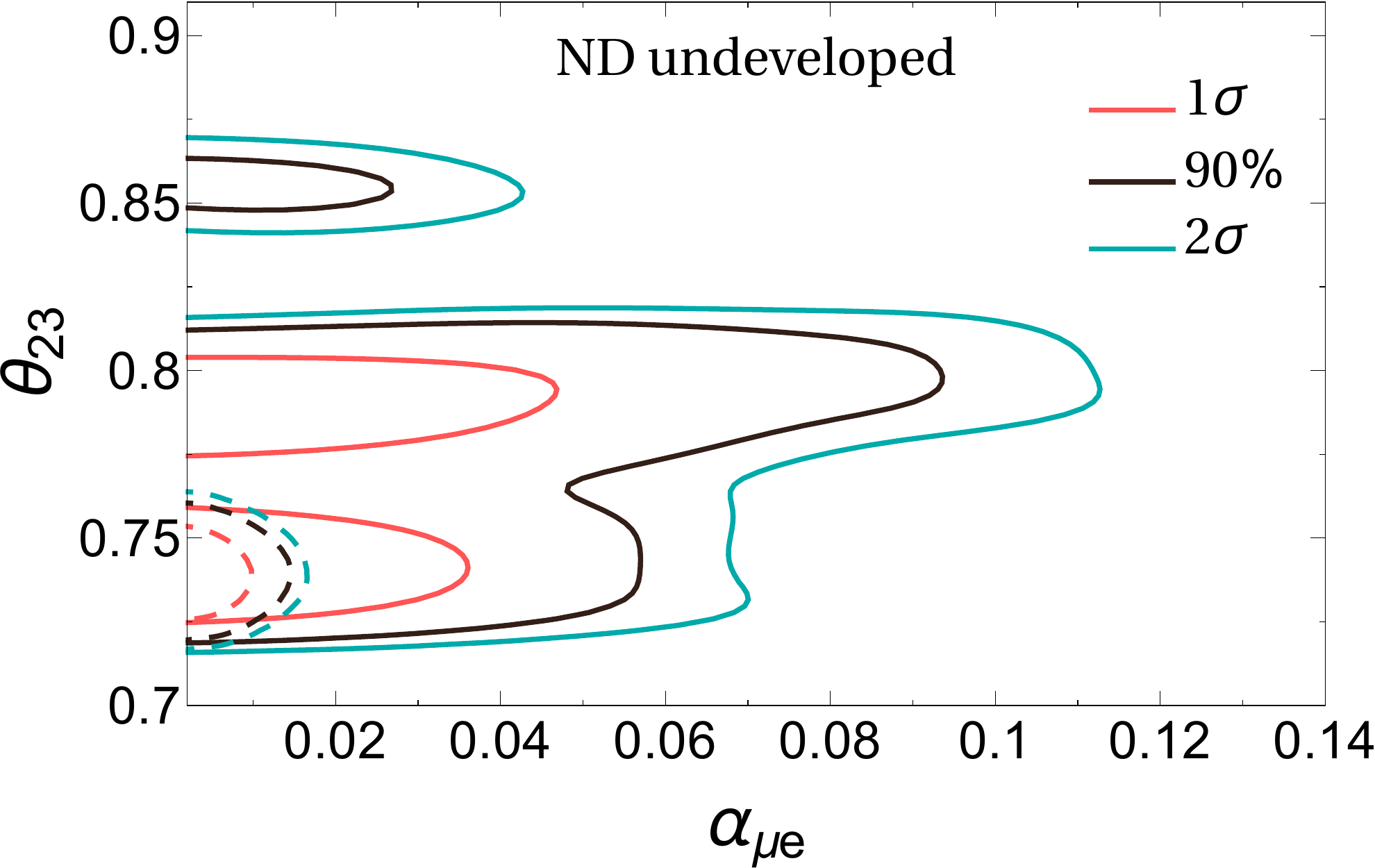} \\
 \includegraphics[width=0.48\columnwidth]{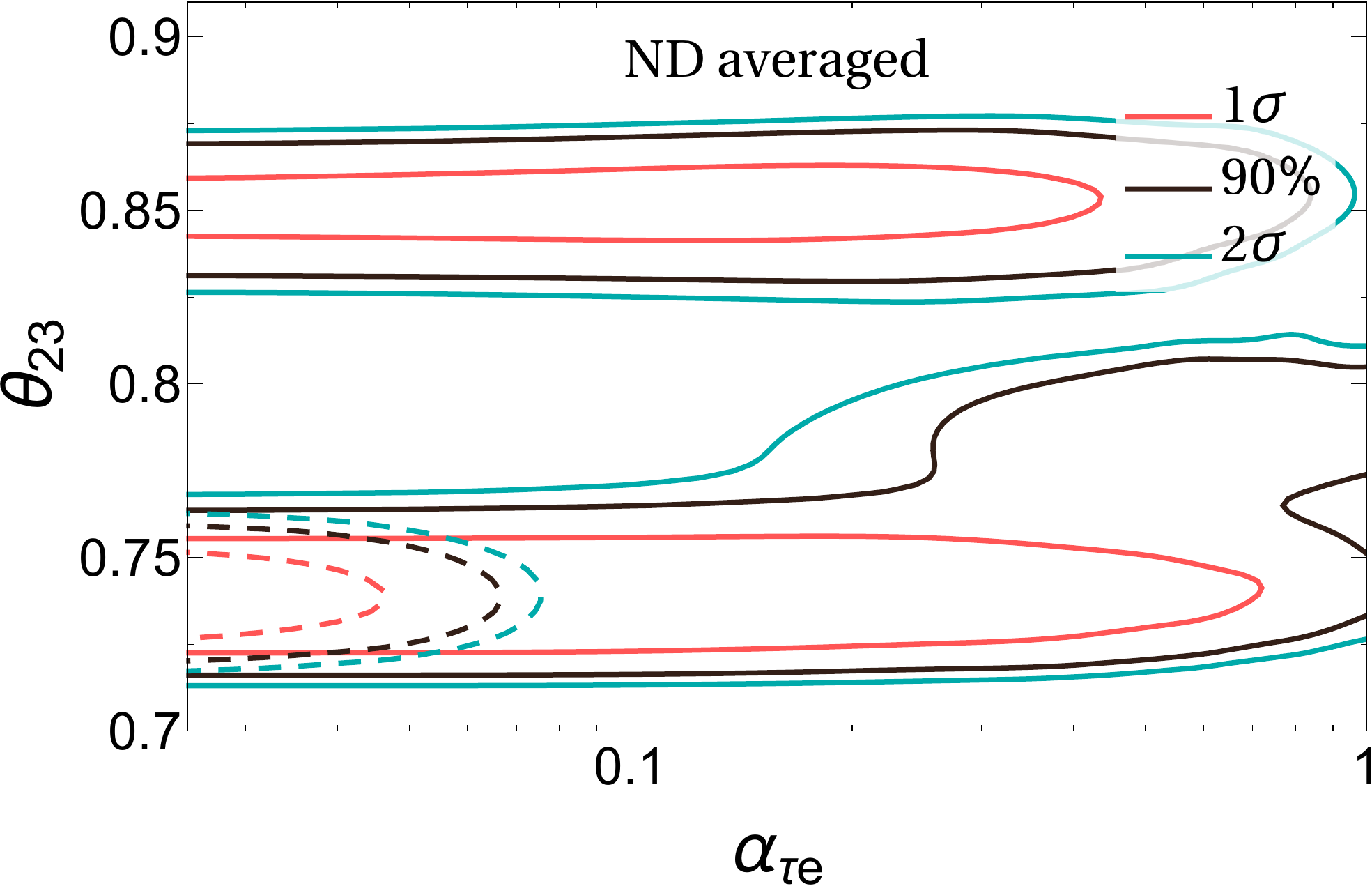} 
 \includegraphics[width=0.48\columnwidth]{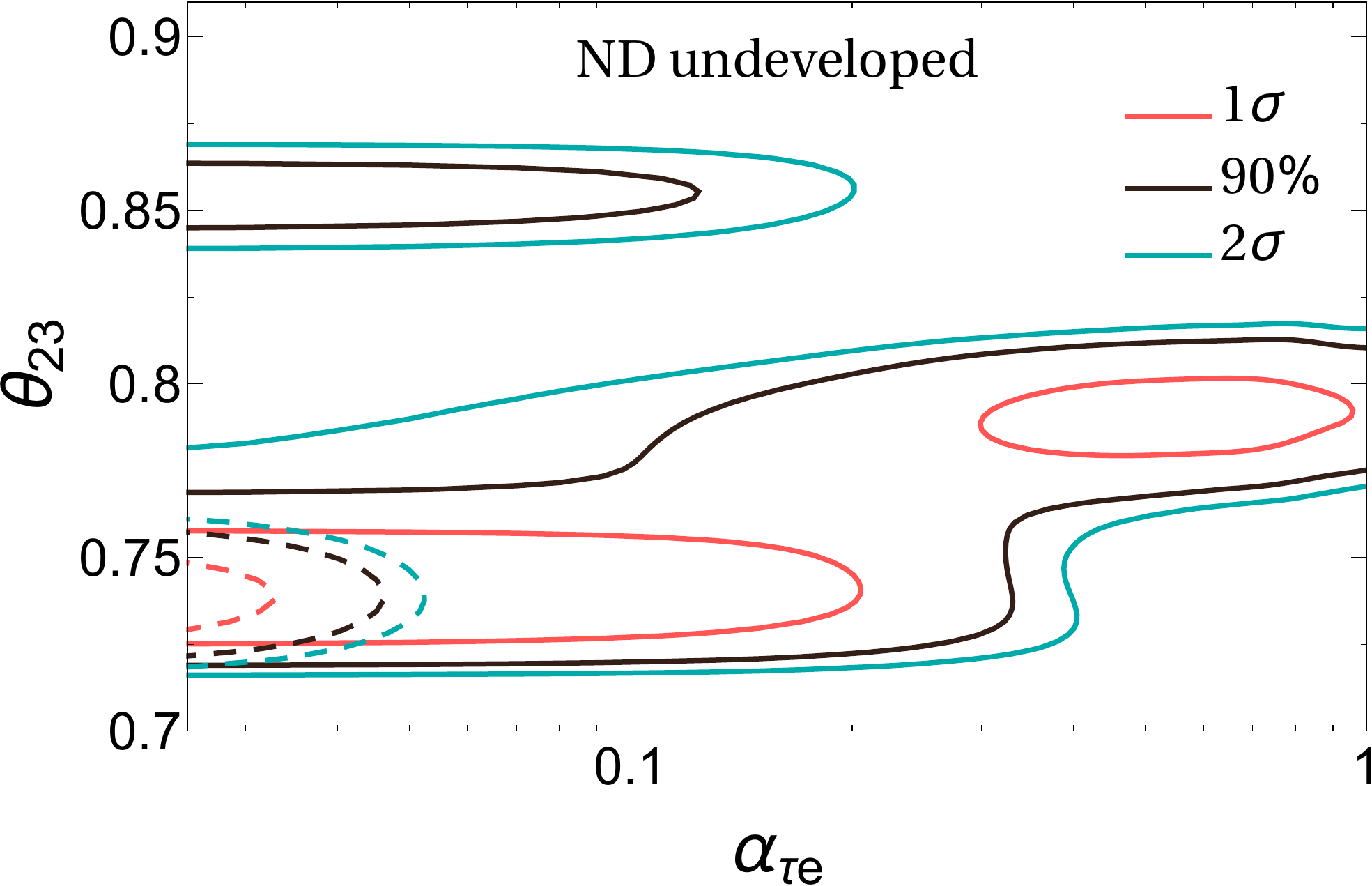}
 \begin{center}
  \caption{ Expected frequentist allowed regions at the $1 \sigma$, $90\%$ and $2 \sigma$ CL for DUNE. All new physics parameters are assumed to be zero so as to obtain the expected sensitivities. The left panels (ND averaged) correspond to the non-unitarity case, or to the sterile case when the light-heavy oscillations are averaged out in the near and far detectors. The right panels (ND undeveloped) give the sensitivity for the sterile case when the light-heavy oscillations have not yet developed in the near detector, but are averaged out in the far. The solid lines correspond to the analysis of DUNE data alone, while the dashed lines include the present constraints on sterile neutrino mixing from the middle and right columns in Tab.~\ref{tab:bounds} for the ND averaged and ND undeveloped scenarios respectively.\label{fig:degs}}
 \end{center}
\end{figure}

Figures~\ref{fig:plots} and~\ref{fig:degs} show the expected sensitivities to the new physics parameters. These have been obtained by assuming that the true values of all $\alpha$ entries are zero to obtain the corresponding expected number of events, and fitting for the corresponding parameters while marginalizing over all other standard and new physics parameters. The resulting frequentist allowed regions are shown at  at $1 \sigma$, $90\%$, and $2 \sigma$~C.L.

The sensitivities obtained for all parameters fall at least one order of magnitude short of the current bounds on the non-unitarity from heavy neutrino scenario presented in Tab.~\ref{tab:bounds}. Thus, the standard three-family oscillations explored at DUNE (and the other present and near-future oscillation facilities) will be free from the possible ambiguities that could otherwise be induced by this type of new physics~\cite{Miranda:2016wdr,Ge:2016xya,Verma:2016nfi,Dutta:2016czj}. While these bounds on non-unitarity are too strong for these effects to be probed at present and near-future facilities a Neutrino Factory~\cite{Geer:1997iz,DeRujula:1998umv} could be precise enough to explore these effects~\cite{FernandezMartinez:2007ms,Antusch:2009pm,Meloni:2009cg}. The situation is slightly different if the results are interpreted in terms of an averaged-out sterile neutrino, since present constraints are weaker in this case. We will therefore focus on this scenario for the rest of our discussion and also study the case in which DUNE data is complemented by our present prior constraints on the sterile neutrino mixing (middle and right columns of Tab.~\ref{tab:bounds} for the ND averaged and undeveloped scenarios respectively), since synergies between the data sets may be present. This case is depicted with dashed lines in Figs.~\ref{fig:plots} and~\ref{fig:degs}. As an example of such synergy, the sensitivity to the real part of $\alpha_{\tau\mu}$ improves for the ND undeveloped scenario through the combination of DUNE data and the present priors with respect to both datasets independently. Indeed, the prior on its own would give the same bound for the real an imaginary parts (as for the ND averaged case in the left panel) and its value roughly corresponds to the constraint obtained for the the imaginary part of $\alpha_{\tau\mu}$, while the sensitivity to the real part does improve through the combination with DUNE.

Another conclusion that can be drawn from Fig.~\ref{fig:plots} is that the sensitivities to the diagonal parameters $\alpha_{ee}$ and $\alpha_{\mu\mu}$ are significantly stronger for the {ND undeveloped} (right panels) as compared to the {ND averaged} scenario (left panels). This was to be expected since the source and detection effects that provide a leading order sensitivity to the diagonal parameters are totally or partially cancelled once the normalization of Eq.~\eqref{eq:experimentalprobability} is included (see Eq.~(\ref{eq:cancellation})). In the disappearance channel both effects cancel in the ratio, while for the appearance channel there is a partial cancellation that only allows the experiment to be sensitive to the combination $\alpha_{ee}-\alpha_{\mu \mu}$. This leads to a pronounced correlation among $\alpha_{ee}$ and $\alpha_{\mu \mu}$, seen in the upper left panel of Fig.~\ref{fig:plots}.

From a phenomenological point of view we observe that, if both near and far detectors are affected by the new physics in the same way (as is the case when the sterile neutrino oscillations are averaged out at both detectors, or in the non-unitarity scenario) their effects are more difficult to observe since they cannot be disentangled from the flux and cross section determination at the near detector. Conversely, in the case in which sterile neutrino oscillations have not yet developed at the near detector but are averaged out at the far, the flux determined by both detectors will have a different normalization. Thus, a strong linear sensitivity to $\alpha_{ee}$ and $\alpha_{\mu\mu}$ is obtained from detector and source effects respectively, although there is no improvement over present constraints.

Fig.~\ref{fig:plots} also shows strong correlations in the middle left panel, involving $\alpha_{\tau \tau}$ and $\alpha_{ee}$. Indeed, sensitivity to $ \alpha_{\tau\tau}$ comes through the matter effects, which only depend on the diagonal entries through their differences $\alpha_{\beta \beta} -\alpha_{\gamma \gamma}$, since a global term of the form of $\alpha_{\gamma\gamma}I$ does not affect neutrino oscillations at leading order in $\alpha$. In these panels we also observe a large difference between the allowed regions for $\alpha_{\tau\tau}$ once prior constraints on the $\alpha$ parameters are included in the analysis, by comparing the solid and dashed lines. This is due to the lifting of degeneracies involving $\theta_{23}$ and the combination $\alpha_{\tau\tau} - \alpha_{\mu\mu}$, and will be discussed in more detail below.

Interesting correlations and degeneracies among the standard and new physics parameters can indeed take place in the averaged-out sterile neutrino scenario~\cite{Berryman:2015nua,Agarwalla:2016xxa,Dutta:2016glq,Agarwalla:2016xlg}. In our results, even though the true values of the $\alpha$ parameters were set to zero, some very interesting correlations and degeneracies among $\theta_{23}$ and the new physics parameters have been recovered. These are shown in Fig.~\ref{fig:degs}, and have been noticed in the context of NSI\footnote{Note the correspondence between NSI, steriles, and non-unitarity presented in Section~\ref{sec:NSI}.} in Refs.~\cite{Coloma:2015kiu,deGouvea:2015ndi,Blennow:2016etl,Agarwalla:2016fkh} (for other works on degeneracies among standard and non-standard parameters in DUNE see \emph{e.g.}, Refs.~\cite{Masud:2016gcl,Masud:2016bvp,Masud:2015xva,Coloma:2016gei}). The first degeneracy appears for the wrong octant of $\theta_{23}$, which would otherwise be correctly determined by the interplay between the appearance and disappearance channels at DUNE (see \emph{e.g.}, Ref.~\cite{DeRomeri:2016qwo}). We have checked that this degeneracy is characterized by non zero values of $\alpha_{\tau e}$ with a non-trivial phase around $\pi$. At the same time, positive values of $\alpha_{\mu \mu} - \alpha_{\tau \tau}$ are slightly preferred. From Ref.~\cite{Agarwalla:2016fkh} this degeneracy was expected for the phase of $\phi_{\tau e} =\arg( \alpha_{\tau e}) \sim \pi$ since $\delta_\mathrm{CP} = -\pi/2$ and strong correlations between these two parameters are required in order to reproduce this degeneracy. Note that this degeneracy is partially lifted in the {ND undeveloped} scenario (right panels). Indeed, the strong sensitivity that this scenario presents to $\alpha_{\mu \mu}$ translates into very stringent bounds that do not allow the preferred positive values of $\alpha_{\mu \mu} - \alpha_{\tau \tau}$ seen in the left panels for the {ND averaged} case since the diagonal elements of $\alpha$ are positive (see Eq.~\eqref{eq:sines}). Upon the inclusion of prior constraints this degeneracy is lifted in both scenarios.

Interestingly, the second degeneracy involves values of $\theta_{23} \sim \pi/4$, so that it could potentially compromise the capabilities of DUNE to determine the maximality of this mixing angle. This degeneracy takes place for $\alpha_{\mu \mu} - \alpha_{\tau \tau} \sim -0.6$, and large values of $\alpha_{\mu e}$ and $\alpha_{\tau e}$ are also needed. Fortunately, present constraints on these parameters are already strong enough to also rule out this possibility (see Tab.~\ref{tab:bounds}), so that a clean determination of the maximality of $\theta_{23}$ should be possible at DUNE. Moreover, when the current bound on $\alpha_{\tau e}$ from the right column in Tab.~\ref{tab:bounds} is added as prior to the simulations, the sensitivity to $\alpha_{\mu e}$ is increased slightly beyond the present prior and the allowed region around $\theta_{23}\sim \pi/4$ is ruled out. This example shows explicitly the complementarity between current constraints and DUNE sensitivities.

All in all, we find that, upon solving the degeneracies through the inclusion of present priors, DUNE's sensitivity would slightly improve upon the present constraints on $\alpha_{\mu e}$ in the ND averaged case as well as the real part of $\alpha_{\tau \mu}$ for the ND undeveloped scenario. While the potential improvement over present bounds is marginal, this also implies that, at the confidence levels studied in this work, the sensitivities to the standard three neutrino oscillations are rather robust and not significantly compromised by the new physics investigated here.

\section{Conclusions}
\label{sec:concl}

The simplest and most natural extension of the Standard Model that can account for our present evidence for 
neutrino masses and mixings is the addition of right-handed neutrinos to the Standard Model (SM) particle content. Gauge and Lorentz 
invariance then imply the possible existence of a Majorana mass for these new particles at a scale to be determined 
by observations. In this work we have studied the impact that two limiting regimes for this new physics scale can 
have in neutrino oscillation experiments. For very high Majorana masses, beyond the kinematic reach of our experiments, 
the imprint of these new degrees of freedom at low energies takes the form of unitarity deviations of the PMNS mixing matrix. In the opposite limit, for small Majorana masses, these extra sterile neutrinos are produced and can participate in neutrino oscillations. However, it should be kept in mind that the neutrino oscillation phenomenology discussed here applies also to other types of new physics that could induce unitarity deviations for the PMNS mixing matrix. This includes any model in which heavy fermions mix with the SM neutrinos or charged leptons, as for instance the type-I/type-III seesaw, Left-Right symmetric models, and models with kinematically accessible sterile neutrinos in the averaged-out regime.

Despite being sourced by different underlying physics, we have seen that, when the sterile neutrino oscillations are averaged out 
(and at leading order in the small heavy-active mixing angles) both limits lead to the same modifications in the 
neutrino oscillation probabilities. Namely, a modification of the interactions in the source and detector which 
implies short-distance effects as well as modified matter effects which, contrary to the standard scenario, also 
involve neutral current interactions. However, the present constraints that apply to these two scenarios are very 
different. Indeed, PMNS non-unitarity is bounded at the per mille level, or even better for some elements, through precision electroweak and flavour observables, while sterile neutrino mixing in the averaged-out regime is allowed at the percent level since it can only be probed via oscillation experiments themselves. Thus, PMNS non-unitarity can have no impact in present or near-future oscillation facilities while sterile neutrino mixing could potentially be discovered by them. 

We have also noted apparently conflicting results depending on the parametrization used to encode these new physics effects. The source of this apparent inconsistency was found to be the different quantities that are commonly identified with the standard PMNS matrix in each parametrization. The conflict was solved by providing a mapping between the two sets of parameters and by identifying the parametrization for which these PMNS parameters correspond to what is determined experimentally.

The role of the near detector was also explored in depth. Indeed, since present and near future oscillation experiments constrain their fluxes and detection cross sections using near detector data it is important to consider if the new physics affects the near and far detector measurements in the same way. If this is the case, the source and detector short-distance effects cancel to a large extent, since there is no additional handle to separate them from flux and cross-section uncertainties. This is always the case in the non-unitarity scenario and when sterile neutrino oscillations are averaged out both at the near and far detectors. Conversely, if sterile neutrino oscillations have not developed yet at the near detector, the determination of the flux and cross section is free from new physics ambiguities and, when compared with the far detector data, a greater sensitivity to the flavour-conserving new physics effects is obtained. This crucial difference is sometimes overlooked in the present literature. 
Finally we also provided a mapping of these new physics effects in the popular non-standard interaction (NSI) formalism.

These effects were exemplified through numerical simulations of the proposed DUNE neutrino oscillation experiment. Our simulations confirm that PMNS non-unitarity is indeed beyond the reach of high precision experiments such as DUNE, but that sterile neutrino oscillations could manifest in several possible interesting ways. Indeed, degeneracies between $\theta_{23}$ and the new physics parameters, previously identified in the context of NSI, have been found in our simulations. These degeneracies could potentially compromise the capability of DUNE to determine the maximality of $\theta_{23}$ as well as its ability to discern its correct octant. We find that current bounds on the new physics parameters are able to lift the degeneracies around $\theta_{23}\sim \pi/4$.

Through these simulations the importance of correctly accounting for the impact of the near detector was made evident. 
Indeed, a very significant increase in the sensitivity to the new physics parameters was found for the case in which the 
near detector is not affected in the same way as the far. This would be the case of sterile neutrino oscillations that 
are undeveloped at the near detector but averaged out at the far. However, the parameter space for this situation to 
take place is rather small (for $\Delta m^2 \sim 0.1-1$~eV$^2$). The most common situation would rather be that in 
which sterile neutrino oscillations are averaged out at both near and far detectors. However, this fact has been usually 
overlooked in previous literature.  

The origin of neutrino masses remains one of our best windows to explore the new physics underlying the open problems 
of the SM. Its simplest extension to accommodate neutrino masses and mixings offers a multitude of phenomenological 
consequences that vary depending of the new physics scale introduced and that should be thoroughly explored by 
future searches. 
In this work, we have explored the impact of these new physics in neutrino oscillation phenomena. We have found that neutrino oscillation facilities are best suited to probe the lightest new physics scales, i.e., kinematically accessible sterile neutrinos.

\section*{Acknowledgements}
We warmly thank Joachim Kopp and Rafael Torres for useful discussions. We also thank Mariam Tortola for her useful input pointing out competitive bounds for the sterile neutrino mixings. We acknowledge support from the EU through grants H2020-MSCA-ITN-2015/674896-Elusives and H2020-MSCA-RISE-2015/690575-InvisiblesPlus. This work was supported by the G{\"o}ran Gustafsson Foundation (MB).  EFM and JHG also acknowledge support from the EU FP7 Marie Curie Actions CIG NeuProbes (PCIG11-GA-2012-321582) and the Spanish MINECO through the ``Ram\'on y Cajal'' programme (RYC2011-07710), the project FPA2012-31880,  through the Centro de Excelencia Severo Ochoa Program under grant SEV-2012-0249 and the HPC-Hydra cluster at IFT.  Fermilab is operated by Fermi Research Alliance, LLC under Contract No. DE-AC02-07CH11359 with the United States Department of Energy. EFM and JHG also warmly thank Fermilab for its hospitality during their InvisiblesPlus secondment where this project was initiated.

\appendix

\section{Current constraints on sterile neutrinos}
\label{app}

\begin{table}[htb!]
\setlength{\tabcolsep}{7pt}
\begin{center}
\renewcommand{\arraystretch}{1.6}
\begin{tabular}{|c|c|c|}
\hline
 & Applicability range & Bound \\
\hline
\multirow{1}{*}{$\alpha_{ee}$} & $\Delta m^2 \gtrsim 4$~eV$^2$ & $2.4\cdot 10^{-2}$\cite{Declais:1994su} \\
& $m >$ EW & $1.3\cdot 10^{-3}$\cite{Fernandez-Martinez:2016lgt} \\
\hline
\multirow{1}{*}{$\alpha_{\mu \mu}$} & $\Delta m^2 \gtrsim 0.1$~eV$^2$ & $2.2\cdot 10^{-2}$\cite{Abe:2014gda} \\
& $m >$ EW & $2.2\cdot 10^{-4}$\cite{Fernandez-Martinez:2016lgt} \\
\hline
\multirow{1}{*}{$\alpha_{\tau \tau}$} & $\Delta m^2 \gtrsim 0.1$~eV$^2$ & $1.0\cdot 10^{-1}$\cite{Abe:2014gda} \\
& $m >$ EW & $1.3\cdot 10^{-3}$\cite{Fernandez-Martinez:2016lgt} \\
\hline
\multirow{4}{*}{$|\alpha_{\mu e}|$} & $\Delta m^2 \gtrsim 4$~eV$^2$ & $3.2\cdot 10^{-2}$ 
\\
&  $\Delta m^2 \gtrsim 10$~eV$^2$ & $2.8\cdot 10^{-2}$\cite{Armbruster:2002mp} \\
&  $\Delta m^2 \gtrsim 100$~eV$^2$ & $2.5\cdot 10^{-2}$\cite{Astier:2003gs} \\
&  $\Delta m^2 \gtrsim 1000$~eV$^2$ & $2.3\cdot 10^{-2}$\cite{Avvakumov:2002jj} \\
& $m >$ EW & $6.8\cdot 10^{-4}$\cite{Fernandez-Martinez:2016lgt} \\
\hline
\multirow{1}{*}{$|\alpha_{\tau e}|$} & $\Delta m^2 \gtrsim 4$~eV$^2$ & $6.9\cdot 10^{-2}$ 
\\
& $m >$ EW & $2.7\cdot 10^{-3}$\cite{Fernandez-Martinez:2016lgt} \\
\hline
\multirow{2}{*}{$|\alpha_{\tau \mu}|$} & $\Delta m^2 \gtrsim 0.1$~eV$^2$ & $6.6\cdot 10^{-2}$ 
\\
&  $\Delta m^2 \gtrsim 100$~eV$^2$ & $1.2\cdot 10^{-2}$\cite{Astier:2001yj} \\
& $m >$ EW & $1.2\cdot 10^{-3}$\cite{Fernandez-Martinez:2016lgt} \\
\hline
\end{tabular}
\caption{Summary of the current experimental constraints in the averaged-out regime, applicable to sterile neutrinos 
above a certain mass range. The bounds on the off-diagonal elements which do not have a reference have been obtained 
indirectly from the bounds on the diagonal elements at that scale, using 
$\alpha_{\alpha \beta} \leq 2 \sqrt{\alpha_{\alpha \alpha} \alpha_{\beta \beta}}$ (see Eq.~\eqref{eq:sines}). \label{tab:bounds2}}
\end{center}
\end{table}


In this Appendix we summarize and explain in more detail the current constraints on sterile neutrinos that arise from oscillation searches in the averaged out regime and thus apply for arbitrarily large values of $\Delta m^2$ as well as those stemming from electroweak and flavour precision observables. Notice that, for the latter, some of the observables only apply above the electroweak scale~\cite{Fernandez-Martinez:2016lgt}. Nevertheless, below this scale, stronger constraints from direct searches are available~\cite{Atre:2009rg,Ruchayskiy:2011aa,Drewes:2015iva}. Regarding the oscillation searches, the validity of these constraints will depend on the particular configuration of the experiment used to derive it, which determines when the averaged-out regime is reached. These constraints together with their range of validity are listed 
in Table~\ref{tab:bounds2}. 

The strongest constraints on the mixing with electrons  ($\alpha_{ee}$) stem from the BUGEY-3 experiment~\cite{Declais:1994su}. At this experiment, oscillations enter the averaged-out regime for $\Delta m^2 \gtrsim 4$~eV$^2$. Recent competitive constraints on this parameter by the Daya Bay experiment~\cite{An:2016luf} tend to dominate for smaller $\Delta m^2$ values and are comparable to the bounds from BUGEY-3~\cite{Declais:1994su} around $\Delta m^2 \gtrsim 0.1$~eV$^2$. In the range $\Delta m^2 \gtrsim 0.1-1$~eV$^2$ the bound oscillates significantly between $3.0\cdot 10^{-3}$ and $1.0\cdot 10^{-2}$: therefore, we quote the latter more conservative bound in the rightmost column of Table~\ref{tab:bounds}.  

Current limits on the $\alpha_{\mu \mu}$ and $\alpha_{\tau \tau}$ elements are dominated by the bounds derived from the SK analysis of atmospheric neutrino oscillations~\cite{Abe:2014gda}. These are derived in the averaged-out regime, which in this case corresponds to $\Delta m^2 \gtrsim 0.1$~eV$^2$. For $\alpha_{\mu \mu}$, MINOS~\cite{MINOS:2016viw} sets stronger constraints for lower values of $\Delta m^2$. Again, these oscillate between $4.4\cdot 10^{-3}$ and $1.4\cdot 10^{-2}$ in the range $\Delta m^2 \gtrsim 0.1-1$~eV$^2$. Thus, we quote the more conservative bound in the rightmost column of Table~\ref{tab:bounds}. Regarding $\alpha_{\tau \tau}$, MINOS~\cite{Adamson:2011ku} has similar constraints to the ones from SK atmospherics. Stronger limits are obtained in the global fit in Ref.~\cite{Collin:2016aqd} but only for $\Delta m^2 = 6$~eV$^2$ and not in the averaged-out limit. 

For the off-diagonal elements, the strongest limit for $\alpha_{\mu e}$ stems from the null results of appearance searches 
by NuTeV~\cite{Avvakumov:2002jj} $|\alpha_{e \mu}|<2.3\cdot10^{-2}$, valid once they enter the averaged-out 
regime for $\Delta m^2 \gtrsim 1000$~eV$^2$. Nevertheless, similar bounds from NOMAD~\cite{Astier:2003gs} 
$|\alpha_{e \mu}|<2.5\cdot10^{-2}$ and KARMEN~\cite{Armbruster:2002mp} $|\alpha_{e \mu}|<2.8\cdot10^{-2}$ apply 
for $\Delta m^2 \gtrsim 100$~eV$^2$ and $\Delta m^2 \gtrsim 10$~eV$^2$ respectively. NOMAD~\cite{Astier:2001yj} 
also gives the most stringent constraints for $\alpha_{\tau \mu}$, valid for $\Delta m^2 \gtrsim 100$~eV$^2$.  
For $\alpha_{\tau e}$, the strongest bounds are derived from those on the diagonal elements through 
$\alpha_{\alpha \beta} \leq 2 \sqrt{\alpha_{\alpha \alpha} \alpha_{\beta \beta}}$ (see Eq.~\eqref{eq:sines}). 
Finally, for very light sterile neutrinos, $\Delta m^2\sim 0.1$~eV$^2$, all the direct constraints on the off-diagonal 
elements from NuTeV, NOMAD and KARMEN fade away. In this case, the strongest bounds are obtained indirectly from the
diagonal elements via $\alpha_{\alpha \beta} \leq 2 \sqrt{\alpha_{\alpha \alpha} \alpha_{\beta \beta}}$.


\begin{thebibliography}{85}
\expandafter\ifx\csname natexlab\endcsname\relax\def\natexlab#1{#1}\fi
\expandafter\ifx\csname bibnamefont\endcsname\relax
  \def\bibnamefont#1{#1}\fi
\expandafter\ifx\csname bibfnamefont\endcsname\relax
  \def\bibfnamefont#1{#1}\fi
\expandafter\ifx\csname citenamefont\endcsname\relax
  \def\citenamefont#1{#1}\fi
\expandafter\ifx\csname url\endcsname\relax
  \def\url#1{\texttt{#1}}\fi
\expandafter\ifx\csname urlprefix\endcsname\relax\def\urlprefix{URL }\fi
\providecommand{\bibinfo}[2]{#2}
\providecommand{\eprint}[2][]{\url{#2}}

\bibitem[{\citenamefont{Fukugita and Yanagida}(1986)}]{Fukugita:1986hr}
\bibinfo{author}{\bibfnamefont{M.}~\bibnamefont{Fukugita}} \bibnamefont{and}
  \bibinfo{author}{\bibfnamefont{T.}~\bibnamefont{Yanagida}},
  \bibinfo{journal}{Phys. Lett.} \textbf{\bibinfo{volume}{B174}},
  \bibinfo{pages}{45} (\bibinfo{year}{1986}).

\bibitem[{\citenamefont{Minkowski}(1977)}]{Minkowski:1977sc}
\bibinfo{author}{\bibfnamefont{P.}~\bibnamefont{Minkowski}},
  \bibinfo{journal}{Phys. Lett.} \textbf{\bibinfo{volume}{B67}},
  \bibinfo{pages}{421} (\bibinfo{year}{1977}).

\bibitem[{\citenamefont{Mohapatra and Senjanovic}(1980)}]{Mohapatra:1979ia}
\bibinfo{author}{\bibfnamefont{R.~N.} \bibnamefont{Mohapatra}}
  \bibnamefont{and}
  \bibinfo{author}{\bibfnamefont{G.}~\bibnamefont{Senjanovic}},
  \bibinfo{journal}{Phys. Rev. Lett.} \textbf{\bibinfo{volume}{44}},
  \bibinfo{pages}{912} (\bibinfo{year}{1980}).

\bibitem[{\citenamefont{Yanagida}(1979)}]{Yanagida:1979as}
\bibinfo{author}{\bibfnamefont{T.}~\bibnamefont{Yanagida}}
  (\bibinfo{year}{1979}), \bibinfo{note}{in Proceedings of the Workshop on the
  Baryon Number of the Universe and Unified Theories, Tsukuba, Japan}.

\bibitem[{\citenamefont{Gell-Mann et~al.}()\citenamefont{Gell-Mann, Ramond, and
  Slansky}}]{GellMann:1980vs}
\bibinfo{author}{\bibfnamefont{M.}~\bibnamefont{Gell-Mann}},
  \bibinfo{author}{\bibfnamefont{P.}~\bibnamefont{Ramond}}, \bibnamefont{and}
  \bibinfo{author}{\bibfnamefont{R.}~\bibnamefont{Slansky}},
  \emph{\bibinfo{title}{{Complex Spinors and Unified Theories}}},
  \bibinfo{note}{{Print-80-0576 (CERN)}}.

\bibitem[{\citenamefont{Casas et~al.}(2004)\citenamefont{Casas, Espinosa, and
  Hidalgo}}]{Casas:2004gh}
\bibinfo{author}{\bibfnamefont{J.~A.} \bibnamefont{Casas}},
  \bibinfo{author}{\bibfnamefont{J.~R.} \bibnamefont{Espinosa}},
  \bibnamefont{and} \bibinfo{author}{\bibfnamefont{I.}~\bibnamefont{Hidalgo}},
  \bibinfo{journal}{JHEP} \textbf{\bibinfo{volume}{11}}, \bibinfo{pages}{057}
  (\bibinfo{year}{2004}), \eprint{hep-ph/0410298}.

\bibitem[{\citenamefont{Mohapatra and Valle}(1986)}]{Mohapatra:1986bd}
\bibinfo{author}{\bibfnamefont{R.}~\bibnamefont{Mohapatra}} \bibnamefont{and}
  \bibinfo{author}{\bibfnamefont{J.}~\bibnamefont{Valle}},
  \bibinfo{journal}{Phys.Rev.} \textbf{\bibinfo{volume}{D34}},
  \bibinfo{pages}{1642} (\bibinfo{year}{1986}).

\bibitem[{\citenamefont{Bernabeu et~al.}(1987)\citenamefont{Bernabeu,
  Santamaria, Vidal, Mendez, and Valle}}]{Bernabeu:1987gr}
\bibinfo{author}{\bibfnamefont{J.}~\bibnamefont{Bernabeu}},
  \bibinfo{author}{\bibfnamefont{A.}~\bibnamefont{Santamaria}},
  \bibinfo{author}{\bibfnamefont{J.}~\bibnamefont{Vidal}},
  \bibinfo{author}{\bibfnamefont{A.}~\bibnamefont{Mendez}}, \bibnamefont{and}
  \bibinfo{author}{\bibfnamefont{J.}~\bibnamefont{Valle}},
  \bibinfo{journal}{Phys.Lett.} \textbf{\bibinfo{volume}{B187}},
  \bibinfo{pages}{303} (\bibinfo{year}{1987}).

\bibitem[{\citenamefont{Branco et~al.}(1989)\citenamefont{Branco, Grimus, and
  Lavoura}}]{Branco:1988ex}
\bibinfo{author}{\bibfnamefont{G.~C.} \bibnamefont{Branco}},
  \bibinfo{author}{\bibfnamefont{W.}~\bibnamefont{Grimus}}, \bibnamefont{and}
  \bibinfo{author}{\bibfnamefont{L.}~\bibnamefont{Lavoura}},
  \bibinfo{journal}{Nucl. Phys.} \textbf{\bibinfo{volume}{B312}},
  \bibinfo{pages}{492} (\bibinfo{year}{1989}).

\bibitem[{\citenamefont{Buchmuller and Wyler}(1990)}]{Buchmuller:1990du}
\bibinfo{author}{\bibfnamefont{W.}~\bibnamefont{Buchmuller}} \bibnamefont{and}
  \bibinfo{author}{\bibfnamefont{D.}~\bibnamefont{Wyler}},
  \bibinfo{journal}{Phys.Lett.} \textbf{\bibinfo{volume}{B249}},
  \bibinfo{pages}{458} (\bibinfo{year}{1990}).

\bibitem[{\citenamefont{Pilaftsis}(1992)}]{Pilaftsis:1991ug}
\bibinfo{author}{\bibfnamefont{A.}~\bibnamefont{Pilaftsis}},
  \bibinfo{journal}{Z. Phys.} \textbf{\bibinfo{volume}{C55}},
  \bibinfo{pages}{275} (\bibinfo{year}{1992}), \eprint{hep-ph/9901206}.

\bibitem[{\citenamefont{Kersten and Smirnov}(2007)}]{Kersten:2007vk}
\bibinfo{author}{\bibfnamefont{J.}~\bibnamefont{Kersten}} \bibnamefont{and}
  \bibinfo{author}{\bibfnamefont{A.~Y.} \bibnamefont{Smirnov}},
  \bibinfo{journal}{Phys.Rev.} \textbf{\bibinfo{volume}{D76}},
  \bibinfo{pages}{073005} (\bibinfo{year}{2007}), \eprint{0705.3221}.

\bibitem[{\citenamefont{Abada et~al.}(2007)\citenamefont{Abada, Biggio, Bonnet,
  Gavela, and Hambye}}]{Abada:2007ux}
\bibinfo{author}{\bibfnamefont{A.}~\bibnamefont{Abada}},
  \bibinfo{author}{\bibfnamefont{C.}~\bibnamefont{Biggio}},
  \bibinfo{author}{\bibfnamefont{F.}~\bibnamefont{Bonnet}},
  \bibinfo{author}{\bibfnamefont{M.~B.} \bibnamefont{Gavela}},
  \bibnamefont{and} \bibinfo{author}{\bibfnamefont{T.}~\bibnamefont{Hambye}},
  \bibinfo{journal}{JHEP} \textbf{\bibinfo{volume}{12}}, \bibinfo{pages}{061}
  (\bibinfo{year}{2007}), \eprint{0707.4058}.

\bibitem[{\citenamefont{Antusch et~al.}(2006)\citenamefont{Antusch, Biggio,
  Fernandez-Martinez, Gavela, and Lopez-Pavon}}]{Antusch:2006vwa}
\bibinfo{author}{\bibfnamefont{S.}~\bibnamefont{Antusch}},
  \bibinfo{author}{\bibfnamefont{C.}~\bibnamefont{Biggio}},
  \bibinfo{author}{\bibfnamefont{E.}~\bibnamefont{Fernandez-Martinez}},
  \bibinfo{author}{\bibfnamefont{M.}~\bibnamefont{Gavela}}, \bibnamefont{and}
  \bibinfo{author}{\bibfnamefont{J.}~\bibnamefont{Lopez-Pavon}},
  \bibinfo{journal}{JHEP} \textbf{\bibinfo{volume}{0610}}, \bibinfo{pages}{084}
  (\bibinfo{year}{2006}), \eprint{hep-ph/0607020}.

\bibitem[{\citenamefont{Aguilar et~al.}(2001)}]{Aguilar:2001ty}
\bibinfo{author}{\bibfnamefont{A.}~\bibnamefont{Aguilar}} \bibnamefont{et~al.}
  (\bibinfo{collaboration}{LSND}), \bibinfo{journal}{Phys. Rev.}
  \textbf{\bibinfo{volume}{D64}}, \bibinfo{pages}{112007}
  (\bibinfo{year}{2001}), \eprint{hep-ex/0104049}.

\bibitem[{\citenamefont{Aguilar-Arevalo
  et~al.}(2013)}]{Aguilar-Arevalo:2013pmq}
\bibinfo{author}{\bibfnamefont{A.~A.} \bibnamefont{Aguilar-Arevalo}}
  \bibnamefont{et~al.} (\bibinfo{collaboration}{MiniBooNE}),
  \bibinfo{journal}{Phys. Rev. Lett.} \textbf{\bibinfo{volume}{110}},
  \bibinfo{pages}{161801} (\bibinfo{year}{2013}), \eprint{1303.2588}.

\bibitem[{\citenamefont{Mention et~al.}(2011)\citenamefont{Mention, Fechner,
  Lasserre, Mueller, Lhuillier et~al.}}]{Mention:2011rk}
\bibinfo{author}{\bibfnamefont{G.}~\bibnamefont{Mention}},
  \bibinfo{author}{\bibfnamefont{M.}~\bibnamefont{Fechner}},
  \bibinfo{author}{\bibfnamefont{T.}~\bibnamefont{Lasserre}},
  \bibinfo{author}{\bibfnamefont{T.}~\bibnamefont{Mueller}},
  \bibinfo{author}{\bibfnamefont{D.}~\bibnamefont{Lhuillier}},
  \bibnamefont{et~al.}, \bibinfo{journal}{Phys.Rev.}
  \textbf{\bibinfo{volume}{D83}}, \bibinfo{pages}{073006}
  (\bibinfo{year}{2011}), \eprint{1101.2755}.

\bibitem[{\citenamefont{Huber}(2011)}]{Huber:2011wv}
\bibinfo{author}{\bibfnamefont{P.}~\bibnamefont{Huber}},
  \bibinfo{journal}{Phys. Rev.} \textbf{\bibinfo{volume}{C84}},
  \bibinfo{pages}{024617} (\bibinfo{year}{2011}), \bibinfo{note}{[Erratum:
  Phys. Rev.C85,029901(2012)]}, \eprint{1106.0687}.

\bibitem[{\citenamefont{Kopp et~al.}(2013)\citenamefont{Kopp, Machado, Maltoni,
  and Schwetz}}]{Kopp:2013vaa}
\bibinfo{author}{\bibfnamefont{J.}~\bibnamefont{Kopp}},
  \bibinfo{author}{\bibfnamefont{P.~A.~N.} \bibnamefont{Machado}},
  \bibinfo{author}{\bibfnamefont{M.}~\bibnamefont{Maltoni}}, \bibnamefont{and}
  \bibinfo{author}{\bibfnamefont{T.}~\bibnamefont{Schwetz}},
  \bibinfo{journal}{JHEP} \textbf{\bibinfo{volume}{05}}, \bibinfo{pages}{050}
  (\bibinfo{year}{2013}), \eprint{1303.3011}.

\bibitem[{\citenamefont{Giunti}(2016)}]{Giunti:2015wnd}
\bibinfo{author}{\bibfnamefont{C.}~\bibnamefont{Giunti}},
  \bibinfo{journal}{Nucl. Phys.} \textbf{\bibinfo{volume}{B908}},
  \bibinfo{pages}{336} (\bibinfo{year}{2016}), \eprint{1512.04758}.

\bibitem[{\citenamefont{Collin et~al.}(2016{\natexlab{a}})\citenamefont{Collin,
  Argüelles, Conrad, and Shaevitz}}]{Collin:2016rao}
\bibinfo{author}{\bibfnamefont{G.~H.} \bibnamefont{Collin}},
  \bibinfo{author}{\bibfnamefont{C.~A.} \bibnamefont{Argüelles}},
  \bibinfo{author}{\bibfnamefont{J.~M.} \bibnamefont{Conrad}},
  \bibnamefont{and} \bibinfo{author}{\bibfnamefont{M.~H.}
  \bibnamefont{Shaevitz}}, \bibinfo{journal}{Nucl. Phys.}
  \textbf{\bibinfo{volume}{B908}}, \bibinfo{pages}{354}
  (\bibinfo{year}{2016}{\natexlab{a}}), \eprint{1602.00671}.
	
	\bibitem{Choubey:2016fpi}
  S.~Choubey and D.~Pramanik,
  Phys.\ Lett.\ B {\bf 764} 135 (2017),
  \eprint{1604.04731}.

\bibitem[{\citenamefont{Antonello et~al.}(2015)}]{Antonello:2015lea}
\bibinfo{author}{\bibfnamefont{M.}~\bibnamefont{Antonello}}
  \bibnamefont{et~al.} (\bibinfo{collaboration}{LAr1-ND, ICARUS-WA104,
  MicroBooNE}) (\bibinfo{year}{2015}), \eprint{1503.01520}.

\bibitem[{\citenamefont{Broncano et~al.}(2003)\citenamefont{Broncano, Gavela,
  and Jenkins}}]{Broncano:2002rw}
\bibinfo{author}{\bibfnamefont{A.}~\bibnamefont{Broncano}},
  \bibinfo{author}{\bibfnamefont{M.~B.} \bibnamefont{Gavela}},
  \bibnamefont{and} \bibinfo{author}{\bibfnamefont{E.~E.}
  \bibnamefont{Jenkins}}, \bibinfo{journal}{Phys. Lett.}
  \textbf{\bibinfo{volume}{B552}}, \bibinfo{pages}{177} (\bibinfo{year}{2003}),
  \eprint{hep-ph/0210271}.

\bibitem[{\citenamefont{Fernandez-Martinez
  et~al.}(2007)\citenamefont{Fernandez-Martinez, Gavela, Lopez-Pavon, and
  Yasuda}}]{FernandezMartinez:2007ms}
\bibinfo{author}{\bibfnamefont{E.}~\bibnamefont{Fernandez-Martinez}},
  \bibinfo{author}{\bibfnamefont{M.~B.} \bibnamefont{Gavela}},
  \bibinfo{author}{\bibfnamefont{J.}~\bibnamefont{Lopez-Pavon}},
  \bibnamefont{and} \bibinfo{author}{\bibfnamefont{O.}~\bibnamefont{Yasuda}},
  \bibinfo{journal}{Phys. Lett.} \textbf{\bibinfo{volume}{B649}},
  \bibinfo{pages}{427} (\bibinfo{year}{2007}), \eprint{hep-ph/0703098}.

\bibitem[{\citenamefont{Xing}(2008)}]{Xing:2007zj}
\bibinfo{author}{\bibfnamefont{Z.-z.} \bibnamefont{Xing}},
  \bibinfo{journal}{Phys. Lett.} \textbf{\bibinfo{volume}{B660}},
  \bibinfo{pages}{515} (\bibinfo{year}{2008}), \eprint{0709.2220}.

\bibitem[{\citenamefont{Xing}(2012)}]{Xing:2011ur}
\bibinfo{author}{\bibfnamefont{Z.-z.} \bibnamefont{Xing}},
  \bibinfo{journal}{Phys. Rev.} \textbf{\bibinfo{volume}{D85}},
  \bibinfo{pages}{013008} (\bibinfo{year}{2012}), \eprint{1110.0083}.

\bibitem[{\citenamefont{Escrihuela et~al.}(2015)\citenamefont{Escrihuela,
  Forero, Miranda, Tortola, and Valle}}]{Escrihuela:2015wra}
\bibinfo{author}{\bibfnamefont{F.~J.} \bibnamefont{Escrihuela}},
  \bibinfo{author}{\bibfnamefont{D.~V.} \bibnamefont{Forero}},
  \bibinfo{author}{\bibfnamefont{O.~G.} \bibnamefont{Miranda}},
  \bibinfo{author}{\bibfnamefont{M.}~\bibnamefont{Tortola}}, \bibnamefont{and}
  \bibinfo{author}{\bibfnamefont{J.~W.~F.} \bibnamefont{Valle}},
  \bibinfo{journal}{Phys. Rev.} \textbf{\bibinfo{volume}{D92}},
  \bibinfo{pages}{053009} (\bibinfo{year}{2015}), \eprint{1503.08879}.

\bibitem[{\citenamefont{Li and Luo}(2016)}]{Li:2015oal}
\bibinfo{author}{\bibfnamefont{Y.-F.} \bibnamefont{Li}} \bibnamefont{and}
  \bibinfo{author}{\bibfnamefont{S.}~\bibnamefont{Luo}},
  \bibinfo{journal}{Phys. Rev.} \textbf{\bibinfo{volume}{D93}},
  \bibinfo{pages}{033008} (\bibinfo{year}{2016}), \eprint{1508.00052}.

\bibitem[{\citenamefont{Antusch
  et~al.}(2009{\natexlab{a}})\citenamefont{Antusch, Blennow,
  Fernandez-Martinez, and Lopez-Pavon}}]{Antusch:2009pm}
\bibinfo{author}{\bibfnamefont{S.}~\bibnamefont{Antusch}},
  \bibinfo{author}{\bibfnamefont{M.}~\bibnamefont{Blennow}},
  \bibinfo{author}{\bibfnamefont{E.}~\bibnamefont{Fernandez-Martinez}},
  \bibnamefont{and}
  \bibinfo{author}{\bibfnamefont{J.}~\bibnamefont{Lopez-Pavon}},
  \bibinfo{journal}{Phys. Rev.} \textbf{\bibinfo{volume}{D80}},
  \bibinfo{pages}{033002} (\bibinfo{year}{2009}{\natexlab{a}}),
  \eprint{0903.3986}.

\bibitem[{\citenamefont{Antusch
  et~al.}(2009{\natexlab{b}})\citenamefont{Antusch, Baumann, and
  Fernandez-Martinez}}]{Antusch:2008tz}
\bibinfo{author}{\bibfnamefont{S.}~\bibnamefont{Antusch}},
  \bibinfo{author}{\bibfnamefont{J.~P.} \bibnamefont{Baumann}},
  \bibnamefont{and}
  \bibinfo{author}{\bibfnamefont{E.}~\bibnamefont{Fernandez-Martinez}},
  \bibinfo{journal}{Nucl.Phys.} \textbf{\bibinfo{volume}{B810}},
  \bibinfo{pages}{369} (\bibinfo{year}{2009}{\natexlab{b}}),
  \eprint{0807.1003}.

\bibitem[{\citenamefont{Shrock}(1980)}]{Shrock:1980vy}
\bibinfo{author}{\bibfnamefont{R.~E.} \bibnamefont{Shrock}},
  \bibinfo{journal}{Phys. Lett.} \textbf{\bibinfo{volume}{B96}},
  \bibinfo{pages}{159} (\bibinfo{year}{1980}).

\bibitem[{\citenamefont{Schechter and Valle}(1980)}]{Schechter:1980gr}
\bibinfo{author}{\bibfnamefont{J.}~\bibnamefont{Schechter}} \bibnamefont{and}
  \bibinfo{author}{\bibfnamefont{J.~W.~F.} \bibnamefont{Valle}},
  \bibinfo{journal}{Phys. Rev.} \textbf{\bibinfo{volume}{D22}},
  \bibinfo{pages}{2227} (\bibinfo{year}{1980}).

\bibitem[{\citenamefont{Shrock}(1981{\natexlab{a}})}]{Shrock:1980ct}
\bibinfo{author}{\bibfnamefont{R.~E.} \bibnamefont{Shrock}},
  \bibinfo{journal}{Phys. Rev.} \textbf{\bibinfo{volume}{D24}},
  \bibinfo{pages}{1232} (\bibinfo{year}{1981}{\natexlab{a}}).

\bibitem[{\citenamefont{Shrock}(1981{\natexlab{b}})}]{Shrock:1981wq}
\bibinfo{author}{\bibfnamefont{R.~E.} \bibnamefont{Shrock}},
  \bibinfo{journal}{Phys. Rev.} \textbf{\bibinfo{volume}{D24}},
  \bibinfo{pages}{1275} (\bibinfo{year}{1981}{\natexlab{b}}).

\bibitem[{\citenamefont{Langacker and London}(1988)}]{Langacker:1988ur}
\bibinfo{author}{\bibfnamefont{P.}~\bibnamefont{Langacker}} \bibnamefont{and}
  \bibinfo{author}{\bibfnamefont{D.}~\bibnamefont{London}},
  \bibinfo{journal}{Phys.Rev.} \textbf{\bibinfo{volume}{D38}},
  \bibinfo{pages}{886} (\bibinfo{year}{1988}).

\bibitem[{\citenamefont{Bilenky and Giunti}(1993)}]{Bilenky:1992wv}
\bibinfo{author}{\bibfnamefont{S.~M.} \bibnamefont{Bilenky}} \bibnamefont{and}
  \bibinfo{author}{\bibfnamefont{C.}~\bibnamefont{Giunti}},
  \bibinfo{journal}{Phys.Lett.} \textbf{\bibinfo{volume}{B300}},
  \bibinfo{pages}{137} (\bibinfo{year}{1993}), \eprint{hep-ph/9211269}.

\bibitem[{\citenamefont{Nardi et~al.}(1994)\citenamefont{Nardi, Roulet, and
  Tommasini}}]{Nardi:1994iv}
\bibinfo{author}{\bibfnamefont{E.}~\bibnamefont{Nardi}},
  \bibinfo{author}{\bibfnamefont{E.}~\bibnamefont{Roulet}}, \bibnamefont{and}
  \bibinfo{author}{\bibfnamefont{D.}~\bibnamefont{Tommasini}},
  \bibinfo{journal}{Phys.Lett.} \textbf{\bibinfo{volume}{B327}},
  \bibinfo{pages}{319} (\bibinfo{year}{1994}), \eprint{hep-ph/9402224}.

\bibitem[{\citenamefont{Tommasini et~al.}(1995)\citenamefont{Tommasini,
  Barenboim, Bernabeu, and Jarlskog}}]{Tommasini:1995ii}
\bibinfo{author}{\bibfnamefont{D.}~\bibnamefont{Tommasini}},
  \bibinfo{author}{\bibfnamefont{G.}~\bibnamefont{Barenboim}},
  \bibinfo{author}{\bibfnamefont{J.}~\bibnamefont{Bernabeu}}, \bibnamefont{and}
  \bibinfo{author}{\bibfnamefont{C.}~\bibnamefont{Jarlskog}},
  \bibinfo{journal}{Nucl.Phys.} \textbf{\bibinfo{volume}{B444}},
  \bibinfo{pages}{451} (\bibinfo{year}{1995}), \eprint{hep-ph/9503228}.

\bibitem[{\citenamefont{Biggio}(2008)}]{Biggio:2008in}
\bibinfo{author}{\bibfnamefont{C.}~\bibnamefont{Biggio}},
  \bibinfo{journal}{Phys. Lett.} \textbf{\bibinfo{volume}{B668}},
  \bibinfo{pages}{378} (\bibinfo{year}{2008}), \eprint{0806.2558}.

\bibitem[{\citenamefont{Alonso et~al.}(2013)\citenamefont{Alonso, Dhen, Gavela,
  and Hambye}}]{Alonso:2012ji}
\bibinfo{author}{\bibfnamefont{R.}~\bibnamefont{Alonso}},
  \bibinfo{author}{\bibfnamefont{M.}~\bibnamefont{Dhen}},
  \bibinfo{author}{\bibfnamefont{M.}~\bibnamefont{Gavela}}, \bibnamefont{and}
  \bibinfo{author}{\bibfnamefont{T.}~\bibnamefont{Hambye}},
  \bibinfo{journal}{JHEP} \textbf{\bibinfo{volume}{1301}}, \bibinfo{pages}{118}
  (\bibinfo{year}{2013}), \eprint{1209.2679}.
	
\bibitem{Akhmedov:2013hec}
  E.~Akhmedov, A.~Kartavtsev, M.~Lindner, L.~Michaels and J.~Smirnov,
  JHEP {\bf 1305} 081 (2013),
\eprint{1302.1872}.

\bibitem[{\citenamefont{Antusch and Fischer}(2014)}]{Antusch:2014woa}
\bibinfo{author}{\bibfnamefont{S.}~\bibnamefont{Antusch}} \bibnamefont{and}
  \bibinfo{author}{\bibfnamefont{O.}~\bibnamefont{Fischer}},
  \bibinfo{journal}{JHEP} \textbf{\bibinfo{volume}{1410}}, \bibinfo{pages}{94}
  (\bibinfo{year}{2014}), \eprint{1407.6607}.

\bibitem[{\citenamefont{Abada and Toma}(2016{\natexlab{a}})}]{Abada:2015trh}
\bibinfo{author}{\bibfnamefont{A.}~\bibnamefont{Abada}} \bibnamefont{and}
  \bibinfo{author}{\bibfnamefont{T.}~\bibnamefont{Toma}},
  \bibinfo{journal}{JHEP} \textbf{\bibinfo{volume}{02}}, \bibinfo{pages}{174}
  (\bibinfo{year}{2016}{\natexlab{a}}), \eprint{1511.03265}.

\bibitem[{\citenamefont{Abada and Toma}(2016{\natexlab{b}})}]{Abada:2016awd}
\bibinfo{author}{\bibfnamefont{A.}~\bibnamefont{Abada}} \bibnamefont{and}
  \bibinfo{author}{\bibfnamefont{T.}~\bibnamefont{Toma}}
  (\bibinfo{year}{2016}{\natexlab{b}}), \eprint{1605.07643}.

\bibitem[{\citenamefont{Fernandez-Martinez
  et~al.}(2016)\citenamefont{Fernandez-Martinez, Hernandez-Garcia, and
  Lopez-Pavon}}]{Fernandez-Martinez:2016lgt}
\bibinfo{author}{\bibfnamefont{E.}~\bibnamefont{Fernandez-Martinez}},
  \bibinfo{author}{\bibfnamefont{J.}~\bibnamefont{Hernandez-Garcia}},
  \bibnamefont{and}
  \bibinfo{author}{\bibfnamefont{J.}~\bibnamefont{Lopez-Pavon}},
  \bibinfo{journal}{JHEP} \textbf{\bibinfo{volume}{08}}, \bibinfo{pages}{033}
  (\bibinfo{year}{2016}), \eprint{1605.08774}.

\bibitem[{\citenamefont{Forero et~al.}(2011)\citenamefont{Forero, Morisi,
  Tortola, and Valle}}]{Forero:2011pc}
\bibinfo{author}{\bibfnamefont{D.~V.} \bibnamefont{Forero}},
  \bibinfo{author}{\bibfnamefont{S.}~\bibnamefont{Morisi}},
  \bibinfo{author}{\bibfnamefont{M.}~\bibnamefont{Tortola}}, \bibnamefont{and}
  \bibinfo{author}{\bibfnamefont{J.~W.~F.} \bibnamefont{Valle}},
  \bibinfo{journal}{JHEP} \textbf{\bibinfo{volume}{09}}, \bibinfo{pages}{142}
  (\bibinfo{year}{2011}), \eprint{1107.6009}.

\bibitem[{\citenamefont{Declais et~al.}(1995)\citenamefont{Declais, Favier,
  Metref, Pessard, Achkar et~al.}}]{Declais:1994su}
\bibinfo{author}{\bibfnamefont{Y.}~\bibnamefont{Declais}},
  \bibinfo{author}{\bibfnamefont{J.}~\bibnamefont{Favier}},
  \bibinfo{author}{\bibfnamefont{A.}~\bibnamefont{Metref}},
  \bibinfo{author}{\bibfnamefont{H.}~\bibnamefont{Pessard}},
  \bibinfo{author}{\bibfnamefont{B.}~\bibnamefont{Achkar}},
  \bibnamefont{et~al.}, \bibinfo{journal}{Nucl.Phys.}
  \textbf{\bibinfo{volume}{B434}}, \bibinfo{pages}{503} (\bibinfo{year}{1995}).

\bibitem[{\citenamefont{Abe et~al.}(2015)}]{Abe:2014gda}
\bibinfo{author}{\bibfnamefont{K.}~\bibnamefont{Abe}} \bibnamefont{et~al.}
  (\bibinfo{collaboration}{Super-Kamiokande}), \bibinfo{journal}{Phys. Rev.}
  \textbf{\bibinfo{volume}{D91}}, \bibinfo{pages}{052019}
  (\bibinfo{year}{2015}), \eprint{1410.2008}.

\bibitem[{\citenamefont{Adamson et~al.}(2016)}]{MINOS:2016viw}
\bibinfo{author}{\bibfnamefont{P.}~\bibnamefont{Adamson}} \bibnamefont{et~al.}
  (\bibinfo{collaboration}{MINOS}), \bibinfo{journal}{Submitted to: Phys. Rev.
  Lett.}  (\bibinfo{year}{2016}), \eprint{1607.01176}.

\bibitem[{\citenamefont{Astier et~al.}(2003)}]{Astier:2003gs}
\bibinfo{author}{\bibfnamefont{P.}~\bibnamefont{Astier}} \bibnamefont{et~al.}
  (\bibinfo{collaboration}{NOMAD}), \bibinfo{journal}{Phys. Lett.}
  \textbf{\bibinfo{volume}{B570}}, \bibinfo{pages}{19} (\bibinfo{year}{2003}),
  \eprint{hep-ex/0306037}.

\bibitem[{\citenamefont{Astier et~al.}(2001)}]{Astier:2001yj}
\bibinfo{author}{\bibfnamefont{P.}~\bibnamefont{Astier}} \bibnamefont{et~al.}
  (\bibinfo{collaboration}{NOMAD}), \bibinfo{journal}{Nucl. Phys.}
  \textbf{\bibinfo{volume}{B611}}, \bibinfo{pages}{3} (\bibinfo{year}{2001}),
  \eprint{hep-ex/0106102}.

\bibitem[{\citenamefont{Atre et~al.}(2009)\citenamefont{Atre, Han, Pascoli, and
  Zhang}}]{Atre:2009rg}
\bibinfo{author}{\bibfnamefont{A.}~\bibnamefont{Atre}},
  \bibinfo{author}{\bibfnamefont{T.}~\bibnamefont{Han}},
  \bibinfo{author}{\bibfnamefont{S.}~\bibnamefont{Pascoli}}, \bibnamefont{and}
  \bibinfo{author}{\bibfnamefont{B.}~\bibnamefont{Zhang}},
  \bibinfo{journal}{JHEP} \textbf{\bibinfo{volume}{05}}, \bibinfo{pages}{030}
  (\bibinfo{year}{2009}), \eprint{0901.3589}.

\bibitem[{\citenamefont{Ruchayskiy and Ivashko}(2012)}]{Ruchayskiy:2011aa}
\bibinfo{author}{\bibfnamefont{O.}~\bibnamefont{Ruchayskiy}} \bibnamefont{and}
  \bibinfo{author}{\bibfnamefont{A.}~\bibnamefont{Ivashko}},
  \bibinfo{journal}{JHEP} \textbf{\bibinfo{volume}{06}}, \bibinfo{pages}{100}
  (\bibinfo{year}{2012}), \eprint{1112.3319}.

\bibitem[{\citenamefont{Drewes and Garbrecht}(2015)}]{Drewes:2015iva}
\bibinfo{author}{\bibfnamefont{M.}~\bibnamefont{Drewes}} \bibnamefont{and}
  \bibinfo{author}{\bibfnamefont{B.}~\bibnamefont{Garbrecht}}
  (\bibinfo{year}{2015}), \eprint{1502.00477}.

\bibitem[{\citenamefont{Parke and Ross-Lonergan}(2015)}]{Parke:2015goa}
\bibinfo{author}{\bibfnamefont{S.}~\bibnamefont{Parke}} \bibnamefont{and}
  \bibinfo{author}{\bibfnamefont{M.}~\bibnamefont{Ross-Lonergan}}
  (\bibinfo{year}{2015}), \eprint{1508.05095}.

\bibitem[{\citenamefont{Dutta et~al.}(2016{\natexlab{a}})\citenamefont{Dutta,
  Gandhi, Kayser, Masud, and Prakash}}]{Dutta:2016glq}
\bibinfo{author}{\bibfnamefont{D.}~\bibnamefont{Dutta}},
  \bibinfo{author}{\bibfnamefont{R.}~\bibnamefont{Gandhi}},
  \bibinfo{author}{\bibfnamefont{B.}~\bibnamefont{Kayser}},
  \bibinfo{author}{\bibfnamefont{M.}~\bibnamefont{Masud}}, \bibnamefont{and}
  \bibinfo{author}{\bibfnamefont{S.}~\bibnamefont{Prakash}}
  (\bibinfo{year}{2016}{\natexlab{a}}), \eprint{1607.02152}.

\bibitem[{\citenamefont{Huber et~al.}(2005)\citenamefont{Huber, Lindner, and
  Winter}}]{Huber:2004ka}
\bibinfo{author}{\bibfnamefont{P.}~\bibnamefont{Huber}},
  \bibinfo{author}{\bibfnamefont{M.}~\bibnamefont{Lindner}}, \bibnamefont{and}
  \bibinfo{author}{\bibfnamefont{W.}~\bibnamefont{Winter}},
  \bibinfo{journal}{Comput. Phys. Commun.} \textbf{\bibinfo{volume}{167}},
  \bibinfo{pages}{195} (\bibinfo{year}{2005}), \eprint{hep-ph/0407333}.

\bibitem[{\citenamefont{Huber et~al.}(2007)\citenamefont{Huber, Kopp, Lindner,
  Rolinec, and Winter}}]{Huber:2007ji}
\bibinfo{author}{\bibfnamefont{P.}~\bibnamefont{Huber}},
  \bibinfo{author}{\bibfnamefont{J.}~\bibnamefont{Kopp}},
  \bibinfo{author}{\bibfnamefont{M.}~\bibnamefont{Lindner}},
  \bibinfo{author}{\bibfnamefont{M.}~\bibnamefont{Rolinec}}, \bibnamefont{and}
  \bibinfo{author}{\bibfnamefont{W.}~\bibnamefont{Winter}},
  \bibinfo{journal}{Comput. Phys. Commun.} \textbf{\bibinfo{volume}{177}},
  \bibinfo{pages}{432} (\bibinfo{year}{2007}), \eprint{hep-ph/0701187}.

\bibitem[{\citenamefont{Alion et~al.}(2016)}]{Alion:2016uaj}
\bibinfo{author}{\bibfnamefont{T.}~\bibnamefont{Alion}} \bibnamefont{et~al.}
  (\bibinfo{collaboration}{DUNE}) (\bibinfo{year}{2016}), \eprint{1606.09550}.

\bibitem[{\citenamefont{Blennow and Fernandez-Martinez}(2010)}]{Blennow:2009pk}
\bibinfo{author}{\bibfnamefont{M.}~\bibnamefont{Blennow}} \bibnamefont{and}
  \bibinfo{author}{\bibfnamefont{E.}~\bibnamefont{Fernandez-Martinez}},
  \bibinfo{journal}{Comput. Phys. Commun.} \textbf{\bibinfo{volume}{181}},
  \bibinfo{pages}{227} (\bibinfo{year}{2010}), \eprint{0903.3985}.

\bibitem[{\citenamefont{Gonzalez-Garcia
  et~al.}(2014)\citenamefont{Gonzalez-Garcia, Maltoni, and
  Schwetz}}]{Gonzalez-Garcia:2014bfa}
\bibinfo{author}{\bibfnamefont{M.}~\bibnamefont{Gonzalez-Garcia}},
  \bibinfo{author}{\bibfnamefont{M.}~\bibnamefont{Maltoni}}, \bibnamefont{and}
  \bibinfo{author}{\bibfnamefont{T.}~\bibnamefont{Schwetz}},
  \bibinfo{journal}{JHEP} \textbf{\bibinfo{volume}{1411}}, \bibinfo{pages}{052}
  (\bibinfo{year}{2014}), \eprint{1409.5439}.

\bibitem[{\citenamefont{Dziewonski and Anderson}(1981)}]{Dziewonski:1981xy}
\bibinfo{author}{\bibfnamefont{A.~M.} \bibnamefont{Dziewonski}}
  \bibnamefont{and} \bibinfo{author}{\bibfnamefont{D.~L.}
  \bibnamefont{Anderson}}, \bibinfo{journal}{Phys. Earth Planet. Interiors}
  \textbf{\bibinfo{volume}{25}}, \bibinfo{pages}{297} (\bibinfo{year}{1981}).

\bibitem[{\citenamefont{Miranda et~al.}(2016)\citenamefont{Miranda, Tortola,
  and Valle}}]{Miranda:2016wdr}
\bibinfo{author}{\bibfnamefont{O.~G.} \bibnamefont{Miranda}},
  \bibinfo{author}{\bibfnamefont{M.}~\bibnamefont{Tortola}}, \bibnamefont{and}
  \bibinfo{author}{\bibfnamefont{J.~W.~F.} \bibnamefont{Valle}}
  (\bibinfo{year}{2016}), \eprint{1604.05690}.

\bibitem[{\citenamefont{Ge et~al.}(2016)\citenamefont{Ge, Pasquini, Tortola,
  and Valle}}]{Ge:2016xya}
\bibinfo{author}{\bibfnamefont{S.-F.} \bibnamefont{Ge}},
  \bibinfo{author}{\bibfnamefont{P.}~\bibnamefont{Pasquini}},
  \bibinfo{author}{\bibfnamefont{M.}~\bibnamefont{Tortola}}, \bibnamefont{and}
  \bibinfo{author}{\bibfnamefont{J.~W.~F.} \bibnamefont{Valle}}
  (\bibinfo{year}{2016}), \eprint{1605.01670}.

\bibitem[{\citenamefont{Verma and Bhardwaj}(2016)}]{Verma:2016nfi}
\bibinfo{author}{\bibfnamefont{S.}~\bibnamefont{Verma}} \bibnamefont{and}
  \bibinfo{author}{\bibfnamefont{S.}~\bibnamefont{Bhardwaj}}
  (\bibinfo{year}{2016}), \eprint{1609.06412}.

\bibitem[{\citenamefont{Dutta et~al.}(2016{\natexlab{b}})\citenamefont{Dutta,
  Ghoshal, and Roy}}]{Dutta:2016czj}
\bibinfo{author}{\bibfnamefont{D.}~\bibnamefont{Dutta}},
  \bibinfo{author}{\bibfnamefont{P.}~\bibnamefont{Ghoshal}}, \bibnamefont{and}
  \bibinfo{author}{\bibfnamefont{S.}~\bibnamefont{Roy}}
  (\bibinfo{year}{2016}{\natexlab{b}}), \eprint{1609.07094}.

\bibitem[{\citenamefont{Geer}(1998)}]{Geer:1997iz}
\bibinfo{author}{\bibfnamefont{S.}~\bibnamefont{Geer}}, \bibinfo{journal}{Phys.
  Rev.} \textbf{\bibinfo{volume}{D57}}, \bibinfo{pages}{6989}
  (\bibinfo{year}{1998}), \bibinfo{note}{[Erratum: Phys.
  Rev.D59,039903(1999)]}, \eprint{hep-ph/9712290}.

\bibitem[{\citenamefont{De~Rujula et~al.}(1999)\citenamefont{De~Rujula, Gavela,
  and Hernandez}}]{DeRujula:1998umv}
\bibinfo{author}{\bibfnamefont{A.}~\bibnamefont{De~Rujula}},
  \bibinfo{author}{\bibfnamefont{M.~B.} \bibnamefont{Gavela}},
  \bibnamefont{and}
  \bibinfo{author}{\bibfnamefont{P.}~\bibnamefont{Hernandez}},
  \bibinfo{journal}{Nucl. Phys.} \textbf{\bibinfo{volume}{B547}},
  \bibinfo{pages}{21} (\bibinfo{year}{1999}), \eprint{hep-ph/9811390}.

\bibitem[{\citenamefont{Meloni et~al.}(2010)\citenamefont{Meloni, Ohlsson,
  Winter, and Zhang}}]{Meloni:2009cg}
\bibinfo{author}{\bibfnamefont{D.}~\bibnamefont{Meloni}},
  \bibinfo{author}{\bibfnamefont{T.}~\bibnamefont{Ohlsson}},
  \bibinfo{author}{\bibfnamefont{W.}~\bibnamefont{Winter}}, \bibnamefont{and}
  \bibinfo{author}{\bibfnamefont{H.}~\bibnamefont{Zhang}},
  \bibinfo{journal}{JHEP} \textbf{\bibinfo{volume}{04}}, \bibinfo{pages}{041}
  (\bibinfo{year}{2010}), \eprint{0912.2735}.

\bibitem[{\citenamefont{Berryman et~al.}(2015)\citenamefont{Berryman,
  de~Gouvêa, Kelly, and Kobach}}]{Berryman:2015nua}
\bibinfo{author}{\bibfnamefont{J.~M.} \bibnamefont{Berryman}},
  \bibinfo{author}{\bibfnamefont{A.}~\bibnamefont{de~Gouvêa}},
  \bibinfo{author}{\bibfnamefont{K.~J.} \bibnamefont{Kelly}}, \bibnamefont{and}
  \bibinfo{author}{\bibfnamefont{A.}~\bibnamefont{Kobach}},
  \bibinfo{journal}{Phys. Rev.} \textbf{\bibinfo{volume}{D92}},
  \bibinfo{pages}{073012} (\bibinfo{year}{2015}), \eprint{1507.03986}.

\bibitem[{\citenamefont{Agarwalla
  et~al.}(2016{\natexlab{a}})\citenamefont{Agarwalla, Chatterjee, and
  Palazzo}}]{Agarwalla:2016xxa}
\bibinfo{author}{\bibfnamefont{S.~K.} \bibnamefont{Agarwalla}},
  \bibinfo{author}{\bibfnamefont{S.~S.} \bibnamefont{Chatterjee}},
  \bibnamefont{and} \bibinfo{author}{\bibfnamefont{A.}~\bibnamefont{Palazzo}},
  \bibinfo{journal}{JHEP} \textbf{\bibinfo{volume}{09}}, \bibinfo{pages}{016}
  (\bibinfo{year}{2016}{\natexlab{a}}), \eprint{1603.03759}.

\bibitem[{\citenamefont{Agarwalla
  et~al.}(2016{\natexlab{b}})\citenamefont{Agarwalla, Chatterjee, and
  Palazzo}}]{Agarwalla:2016xlg}
\bibinfo{author}{\bibfnamefont{S.~K.} \bibnamefont{Agarwalla}},
  \bibinfo{author}{\bibfnamefont{S.~S.} \bibnamefont{Chatterjee}},
  \bibnamefont{and} \bibinfo{author}{\bibfnamefont{A.}~\bibnamefont{Palazzo}}
  (\bibinfo{year}{2016}{\natexlab{b}}), \eprint{1605.04299}.

\bibitem[{\citenamefont{Coloma}(2016)}]{Coloma:2015kiu}
\bibinfo{author}{\bibfnamefont{P.}~\bibnamefont{Coloma}},
  \bibinfo{journal}{JHEP} \textbf{\bibinfo{volume}{03}}, \bibinfo{pages}{016}
  (\bibinfo{year}{2016}), \eprint{1511.06357}.

\bibitem[{\citenamefont{de~Gouvêa and Kelly}(2016)}]{deGouvea:2015ndi}
\bibinfo{author}{\bibfnamefont{A.}~\bibnamefont{de~Gouvêa}} \bibnamefont{and}
  \bibinfo{author}{\bibfnamefont{K.~J.} \bibnamefont{Kelly}},
  \bibinfo{journal}{Nucl. Phys.} \textbf{\bibinfo{volume}{B908}},
  \bibinfo{pages}{318} (\bibinfo{year}{2016}), \eprint{1511.05562}.

\bibitem[{\citenamefont{Blennow et~al.}(2016)\citenamefont{Blennow, Choubey,
  Ohlsson, Pramanik, and Raut}}]{Blennow:2016etl}
\bibinfo{author}{\bibfnamefont{M.}~\bibnamefont{Blennow}},
  \bibinfo{author}{\bibfnamefont{S.}~\bibnamefont{Choubey}},
  \bibinfo{author}{\bibfnamefont{T.}~\bibnamefont{Ohlsson}},
  \bibinfo{author}{\bibfnamefont{D.}~\bibnamefont{Pramanik}}, \bibnamefont{and}
  \bibinfo{author}{\bibfnamefont{S.~K.} \bibnamefont{Raut}},
  \bibinfo{journal}{JHEP} \textbf{\bibinfo{volume}{08}}, \bibinfo{pages}{090}
  (\bibinfo{year}{2016}), \eprint{1606.08851}.

\bibitem[{\citenamefont{Agarwalla
  et~al.}(2016{\natexlab{c}})\citenamefont{Agarwalla, Chatterjee, and
  Palazzo}}]{Agarwalla:2016fkh}
\bibinfo{author}{\bibfnamefont{S.~K.} \bibnamefont{Agarwalla}},
  \bibinfo{author}{\bibfnamefont{S.~S.} \bibnamefont{Chatterjee}},
  \bibnamefont{and} \bibinfo{author}{\bibfnamefont{A.}~\bibnamefont{Palazzo}}
  (\bibinfo{year}{2016}{\natexlab{c}}), \eprint{1607.01745}.

\bibitem[{\citenamefont{Masud and Mehta}(2016{\natexlab{a}})}]{Masud:2016gcl}
\bibinfo{author}{\bibfnamefont{M.}~\bibnamefont{Masud}} \bibnamefont{and}
  \bibinfo{author}{\bibfnamefont{P.}~\bibnamefont{Mehta}},
  \bibinfo{journal}{Phys. Rev.} \textbf{\bibinfo{volume}{D94}},
  \bibinfo{pages}{053007} (\bibinfo{year}{2016}{\natexlab{a}}),
  \eprint{1606.05662}.

\bibitem[{\citenamefont{Masud and Mehta}(2016{\natexlab{b}})}]{Masud:2016bvp}
\bibinfo{author}{\bibfnamefont{M.}~\bibnamefont{Masud}} \bibnamefont{and}
  \bibinfo{author}{\bibfnamefont{P.}~\bibnamefont{Mehta}},
  \bibinfo{journal}{Phys. Rev.} \textbf{\bibinfo{volume}{D94}},
  \bibinfo{pages}{013014} (\bibinfo{year}{2016}{\natexlab{b}}),
  \eprint{1603.01380}.

\bibitem[{\citenamefont{Masud et~al.}(2016)\citenamefont{Masud, Chatterjee, and
  Mehta}}]{Masud:2015xva}
\bibinfo{author}{\bibfnamefont{M.}~\bibnamefont{Masud}},
  \bibinfo{author}{\bibfnamefont{A.}~\bibnamefont{Chatterjee}},
  \bibnamefont{and} \bibinfo{author}{\bibfnamefont{P.}~\bibnamefont{Mehta}},
  \bibinfo{journal}{J. Phys.} \textbf{\bibinfo{volume}{G43}},
  \bibinfo{pages}{095005} (\bibinfo{year}{2016}), \eprint{1510.08261}.

\bibitem[{\citenamefont{Coloma and Schwetz}(2016)}]{Coloma:2016gei}
\bibinfo{author}{\bibfnamefont{P.}~\bibnamefont{Coloma}} \bibnamefont{and}
  \bibinfo{author}{\bibfnamefont{T.}~\bibnamefont{Schwetz}},
  \bibinfo{journal}{Phys. Rev.} \textbf{\bibinfo{volume}{D94}},
  \bibinfo{pages}{055005} (\bibinfo{year}{2016}), \eprint{1604.05772}.

\bibitem[{\citenamefont{De~Romeri et~al.}(2016)\citenamefont{De~Romeri,
  Fernandez-Martinez, and Sorel}}]{DeRomeri:2016qwo}
\bibinfo{author}{\bibfnamefont{V.}~\bibnamefont{De~Romeri}},
  \bibinfo{author}{\bibfnamefont{E.}~\bibnamefont{Fernandez-Martinez}},
  \bibnamefont{and} \bibinfo{author}{\bibfnamefont{M.}~\bibnamefont{Sorel}}
  (\bibinfo{year}{2016}), \eprint{1607.00293}.

\bibitem[{\citenamefont{Armbruster et~al.}(2002)}]{Armbruster:2002mp}
\bibinfo{author}{\bibfnamefont{B.}~\bibnamefont{Armbruster}}
  \bibnamefont{et~al.} (\bibinfo{collaboration}{KARMEN}),
  \bibinfo{journal}{Phys. Rev.} \textbf{\bibinfo{volume}{D65}},
  \bibinfo{pages}{112001} (\bibinfo{year}{2002}), \eprint{hep-ex/0203021}.

\bibitem[{\citenamefont{Avvakumov et~al.}(2002)}]{Avvakumov:2002jj}
\bibinfo{author}{\bibfnamefont{S.}~\bibnamefont{Avvakumov}}
  \bibnamefont{et~al.} (\bibinfo{collaboration}{NuTeV}),
  \bibinfo{journal}{Phys. Rev. Lett.} \textbf{\bibinfo{volume}{89}},
  \bibinfo{pages}{011804} (\bibinfo{year}{2002}), \eprint{hep-ex/0203018}.

\bibitem[{\citenamefont{An et~al.}(2016)}]{An:2016luf}
\bibinfo{author}{\bibfnamefont{F.~P.} \bibnamefont{An}} \bibnamefont{et~al.}
  (\bibinfo{collaboration}{Daya Bay}), \bibinfo{journal}{Phys. Rev. Lett.}
  \textbf{\bibinfo{volume}{117}}, \bibinfo{pages}{151802}
  (\bibinfo{year}{2016}), \eprint{1607.01174}.

\bibitem[{\citenamefont{Adamson et~al.}(2011)}]{Adamson:2011ku}
\bibinfo{author}{\bibfnamefont{P.}~\bibnamefont{Adamson}} \bibnamefont{et~al.}
  (\bibinfo{collaboration}{MINOS}), \bibinfo{journal}{Phys. Rev. Lett.}
  \textbf{\bibinfo{volume}{107}}, \bibinfo{pages}{011802}
  (\bibinfo{year}{2011}), \eprint{1104.3922}.

\bibitem[{\citenamefont{Collin et~al.}(2016{\natexlab{b}})\citenamefont{Collin,
  Argüelles, Conrad, and Shaevitz}}]{Collin:2016aqd}
\bibinfo{author}{\bibfnamefont{G.~H.} \bibnamefont{Collin}},
  \bibinfo{author}{\bibfnamefont{C.~A.} \bibnamefont{Argüelles}},
  \bibinfo{author}{\bibfnamefont{J.~M.} \bibnamefont{Conrad}},
  \bibnamefont{and} \bibinfo{author}{\bibfnamefont{M.~H.}
  \bibnamefont{Shaevitz}} (\bibinfo{year}{2016}{\natexlab{b}}),
  \eprint{1607.00011}.

\end{thebibliography}
\end{document}